\documentclass[10pt,journal,compsoc]{IEEEtran}
\ifCLASSOPTIONcompsoc
  \usepackage[nocompress]{cite}
\else
  \usepackage{cite}
\fi
\ifCLASSINFOpdf
\else
\fi
\usepackage[square,sort,comma,numbers]{natbib} 
\usepackage{listings}
\usepackage{subcaption}
\usepackage{graphicx}

\usepackage{amsmath}

\usepackage{xcolor}

\definecolor{diffstart}{RGB}{211,211,211}
\definecolor{diffincl}{RGB}{102,205,170}
\definecolor{diffrem}{RGB}{255,69,0}
\definecolor{bostonuniversityred}{rgb}{0.8, 0.0, 0.0}
\definecolor{darkgreen}{rgb}{0.0, 0.2, 0.13}
\definecolor{yellow-green}{rgb}{0.6, 0.8, 0.2}
\definecolor{guppiegreen}{rgb}{0.0, 1.0, 0.5}
\definecolor{pastelmagenta}{rgb}{0.96, 0.6, 0.76}
\definecolor{tearose(rose)}{rgb}{0.96, 0.76, 0.76}

\usepackage{listings}
\lstdefinelanguage{diff}{
	basicstyle=\ttfamily\small,
	morecomment=[f][\color{diffstart}]{@@},
	morecomment=[f][\color{diffincl}]{+\ },
	morecomment=[f][\color{diffrem}]{-\ },
}

\makeatletter
\newif\if@restonecol
\makeatother

\usepackage[linesnumbered,ruled,vlined]{algorithm2e}
\usepackage{algpseudocode,algorithmicx}
\usepackage{amsmath}

\definecolor{dkgreen1}{rgb}{0,0.6,0}
\definecolor{gray1}{rgb}{0.5,0.5,0.5}
\definecolor{mauve1}{rgb}{0.58,0,0.82}
\usepackage{listings}

\usepackage{booktabs}
\usepackage{verbatim}

\usepackage{siunitx}

\usepackage{stfloats}
\usepackage{textcomp}
\usepackage{booktabs, tabularx}
\usepackage{rotating}
\usepackage{xcolor}
\usepackage{cite}

\usepackage{fancybox}
 \usepackage{multirow}

\usepackage{hyperref}

\usepackage{url}
\usepackage{hyperref}

\usepackage{caption}
\usepackage{subcaption}

\usepackage{balance}
\usepackage{footnote}

\usepackage{array}

\usepackage{adjustbox}
\usepackage{threeparttable}

\newcommand{\Hilight}{\makebox[0pt][l]{\color{cyan}\rule[-2pt]{1\linewidth}{8pt}}}

\definecolor{darkgreen}{RGB}{106, 168, 79}
\newcommand{\revised}[1]{\textcolor{black}{#1}}
\newcommand{\junjie}[1]{{\textcolor{black}{#1}}}
\newcommand{\minor}[1]{{\textcolor{black}{#1}}}

\newcommand{\toolS}{StaticTracker\space}
\newcommand{\tool}{StaticTracker}

\definecolor{dkgreen}{rgb}{0,0.6,0}
\definecolor{gray}{rgb}{0.5,0.5,0.5}
\definecolor{mauve}{rgb}{0.58,0,0.82}

\usepackage{array}
\newcolumntype{$}{>{\global\let\currentrowstyle\relax}}
\newcolumntype{^}{>{\currentrowstyle}}

\newcommand{\rqbox}[1]{
\begin{center}
\vspace{-0.2cm}
\cornersize{.1} 
\setlength{\fboxsep}{7pt}
\ovalbox{\begin{minipage}{3.3in}
{\em #1}
\end{minipage}}
\vspace{-0.2cm}

\end{center}}

\definecolor{pgrey}{rgb}{0.46,0.45,0.48}

\newcommand{\RQOne}{Is the SOTA approach good at tracking the evolution of static code warnings?}
\newcommand{\RQTwo}{What are the limitations of the SOTA approach?}
\newcommand{\RQThree}{What is the performance of our proposed approach \tool?}
\newcommand{\pa}[1]{\noindent\textbf{#1}}

\begin{document}
%
\title{Tracking the Evolution of Static Code Warnings: the State-of-the-Art and a Better Approach}
%
%
%
%

\author{}
\IEEEtitleabstractindextext{%
\begin{abstract}
Static bug detection tools help developers detect problems in the code, including bad programming practices and potential defects. 
Recent efforts to integrate static bug detectors in modern software development workflows, such as in code review and continuous integration, are shown to better motivate developers to fix the reported warnings on the fly. A proper mechanism to track the evolution of the reported warnings can better support such integration. Moreover, tracking the static code warnings will benefit many downstream software engineering tasks, such as learning the fix patterns for automated program repair, and learning which warnings are of more interest, so they can be prioritized automatically. \revised{In addition, the utilization of tracking tools enables developers to concentrate on the most recent and actionable static warnings rather than being overwhelmed by the thousands of warnings from the entire project. This, in turn, enhances the utilization of static analysis tools.} Hence, precisely tracking the warnings by static bug detectors is critical to improving the utilization of static bug detectors further.

In this paper, we study the effectiveness of the state-of-the-art (SOTA) solution in tracking static code warnings and propose a better solution based on our analysis of the insufficiency of the SOTA solution. In particular, we examined over 2,000 commits in four large-scale open-source systems (i.e., JClouds, Kafka, Spring-boot, and Guava) and crafted a dataset of 3,451 static code warnings by two static bug detectors (i.e., Spotbugs and PMD). We manually uncovered the ground-truth evolution status of the static warnings: persistent, \minor{removed\textsubscript{fix}}, \minor{removed\textsubscript{non-fix}} and newly-introduced. Upon manual analysis, we identified the main reasons behind the insufficiency of the SOTA solution. Furthermore, we propose \toolS to track static warnings over software development history.
Our evaluation shows that \toolS significantly improves the tracking precision, i.e., \minor{from 64.4\% to 90.3\%} for the evolution statuses combined (\minor{removed\textsubscript{fix}}, \minor{removed\textsubscript{non-fix}} and newly-introduced). 
\end{abstract}

\begin{IEEEkeywords}
static analysis, code refactoring, software evolution, empirical study.
\end{IEEEkeywords}}

\author{Junjie~Li,~\IEEEmembership{Student~Member,~IEEE,}
	and~Jinqiu~Yang,~\IEEEmembership{Member,~IEEE}
	\IEEEcompsocitemizethanks{\IEEEcompsocthanksitem J. Li, and J. Yang are with the Department
		of Computer Science and Software Engineering, Concordia University, Montreal,
		Quebec, Canada.\protect\\
		E-mail: l\_unjie, jinqiuy@encs.concordia.ca}
		}
\maketitle

\IEEEdisplaynontitleabstractindextext

%
\IEEEpeerreviewmaketitle

\IEEEraisesectionheading{\section{Introduction}}
\IEEEPARstart{S}{tatic} bug detection tools have been widely applied in practice to detect potential defects in software. To name a few, both Google and Facebook adopt static bug detectors in their large codebases on a daily basis~\citep{google_static}. However, static bug detectors are known to be underutilized due to various reasons. 
First, static bug detectors report an overwhelming number of warnings, which may be far beyond what resources are allowed to resolve. For example, Spotbugs~\citep{Spotbugs19}, i.e., the spiritual successor of \textit{Findbugs}, reports thousands of or more static code warnings in one version of \textit{JClouds}. 
Second, static bug detectors are known to detect many false positive warnings. The existence of a large number of false positives discourages developers from actively working on resolving the reported warnings. As a result, a significant portion of static code warnings remain unresolved by developers and can hinder software quality.

There have been efforts from a variety of directions to improve the utilization of static bug detection tools, e.g., prioritizing and recommending actionable static warnings and identifying false positive warnings. 
For example, researchers have been working on techniques to identify actionable warnings and reduce false static code warnings, such as recommending actionable warnings by learning from past development history~\citep{quinn, fix_warning_first}. 
On the other hand, recent studies show that by better integrating static bug detectors in software development workflows, such as code review and continuous integration, developers demonstrated a higher response rate in resolving the reported static warnings~\citep{google_static, tricoder}. Developers are presented with much fewer warnings, which are introduced by a new commit, and encounter fewer context switch problems in fixing the warnings.

Making static bug detectors more frequent in workflows such as code review requires proper management of the evolving static code warnings. 
Such proper management is not straightforward. \revised{One way is to adapt differential static analysis to only analyze modified code files~\citep{do2017just}, yet to achieve satisfactory performance.} However, it requires algorithm innovation and non-trivial engineering effort for every static bug-finder.   

Alternatively, we advocate for management that tracks the evolution of static code warnings in the commit history, i.e., shows the \textit{diffs} of the static code warnings from two consecutive software revisions. 
\minor{Tracking the evolution of static code warnings reveals that given a commit, which warnings remain unresolved (i.e., \textit{persistent}) by developers, which warnings are fixed (i.e., \textit{removed\textsubscript{fix}}) by developers, which warnings are removed due to code deletions (i.e., \textit{removed\textsubscript{non-fix}}), and which warnings are \textit{newly-introduced} in the commit.} 
\revised{The analysis of \minor{removed} and newly-introduced warnings helps developers discern trends in warning changes over time. 
Identifying fixed warnings can benefit downstream software engineering research, such as automatic program repair and mining fix patterns of static warnings.
}
\revised{The presentation of the \minor{four} warning types, i.e., \minor{removed\textsubscript{fix}}, \minor{removed\textsubscript{non-fix}}, persistent, and newly-introduced, provides a comprehensive understanding of the status of warnings across consecutive revisions. Developers may be more motivated to utilize static bug detectors if they are provided with a list of newly introduced static code warnings instead of thousands of persistent warnings that developers showed less interest in. In addition, developers could pay attention to the warnings that have been recently resolved and may be motivated to fix similar warnings.
}

More importantly, effective management of static code warnings will benefit many downstream software engineering tasks. To name a few, researchers have been crawling past fixes of static code warnings to provide fix suggestions for new warnings~\citep{Fixpattern17, phoenix}, which has been shown can further improve the utilization of static bug detectors. Furthermore, such concluded fix patterns are shown to be effective in automated program repair techniques~\citep{avatar}.

To this end, there has been little effort to systematically review existing solutions to track the evolution of static code warnings and accordingly propose better solutions. Prior studies rely on simple heuristics to track the static code warnings~\citep{violation_and_fault, fix_warning_first}, i.e., two warnings are identical if they are of the same warning type, in the same file, etc. 
Avgustinov et al.~\citep{Tracking15} present an algorithm that combines various types of information of one warning, compares two warnings in layers, and eventually establishes mappings between two sets of warnings from two software revisions. This algorithm is adopted by recent automated program techniques, and in this study, we refer to it as the state-of-the-art (SOTA) solution. For example, Liu et al.~\citep{Fixpattern17} adapt the SOTA solution to identify warning-fixing commits in software repositories for automated program repair. However, it remains unknown how accurate the SOTA solution is in tracking the static code warnings. An unacceptable performance of the SOTA solution in tracking static code warnings has subsequent negative impacts on the downstream tasks.  

Hence, to foster future research in static code warnings, in this paper, we examine the performance of the SOTA solution in tracking static code warnings and propose a better approach \toolS after analyzing the insufficiency of the SOTA solution. \junjie{In our study, we found that the SOTA approach has limitations in matching the warnings involved in code refactoring, code shift, and volatile metadata due to the use of an anonymous class or method. In light of these limitations, we propose \toolS with three key improvements (1) the detection of code refactoring (2) matching warning pairs using the Hungarian algorithm, and (3) the handling of volatile metadata of static warnings.  
\minor{In addition, \toolS is designed to differentiate between the warnings that are fixed by developers and the ones removed due to non-fix evolution. In comparison, the SOTA solution does not distill the two different categories, i.e., removed\textsubscript{fix} and removed\textsubscript{non-fix}.} 
}

\begin{figure}[hpt]
	\centering
	\scalebox{1.0}{
		\includegraphics[width=\linewidth]{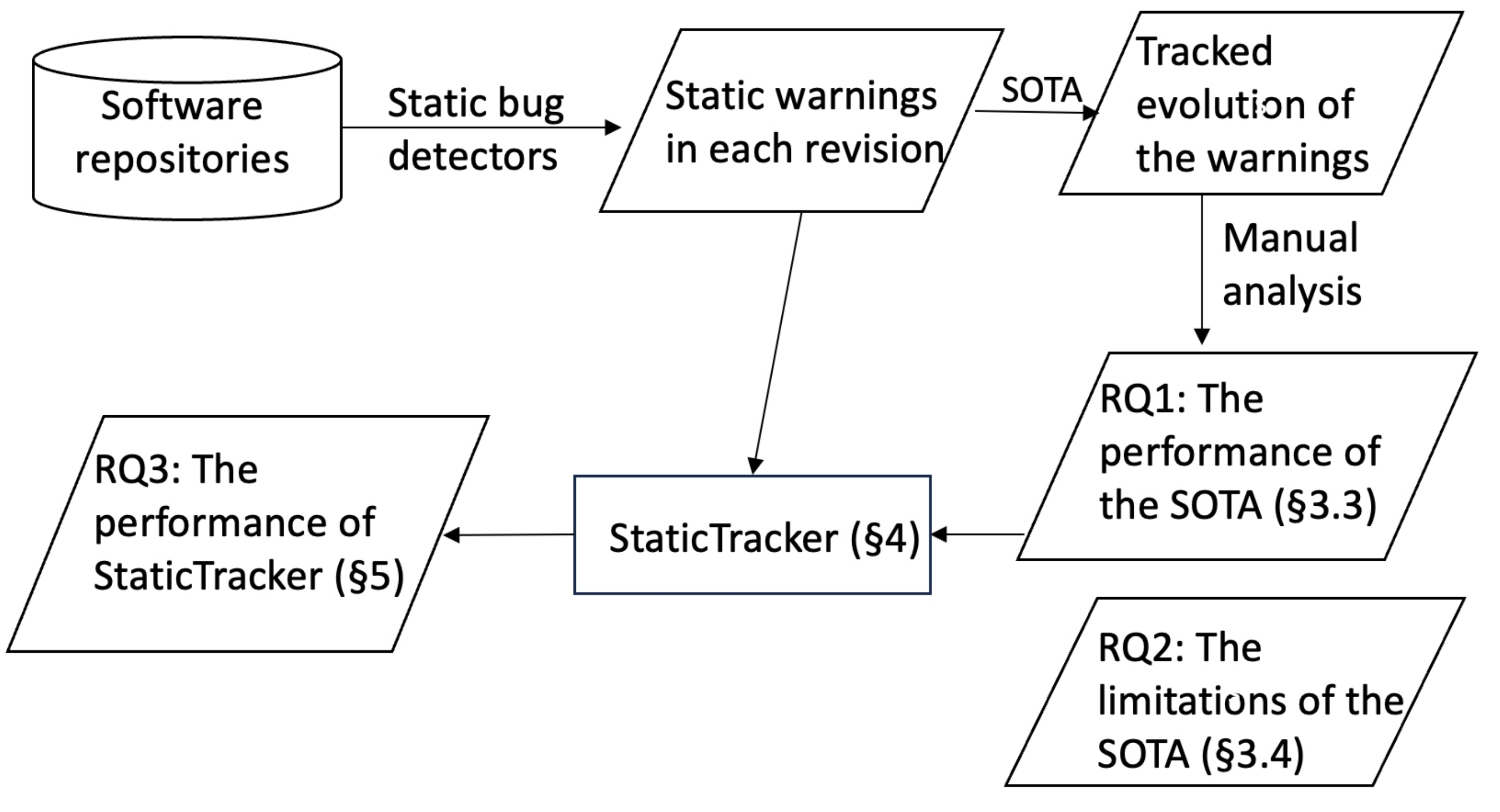}
	}
	\caption{\revised{An overview of our study.}}
	\label{fig:overviewofStudy}
\end{figure}

Figure~\ref{fig:overviewofStudy} shows an overview of this study. In this work, We answer the following research questions:  

\textbf{RQ1} \RQOne

\textbf{RQ2} \RQTwo
 
\revised{\textbf{RQ3} \RQThree}


To answer \textbf{RQ1} and \textbf{RQ2}, we crafted a dataset of static code warning and their evolution. In particular, we took statistically significant samples of the reported static code warnings from the entire development history of four projects (i.e., \textit{JClouds}, \textit{Kafka}, \textit{Guava}, and \textit{Springboot}), and performed manual analysis to label whether each sampled static code warning is \textit{persistent}, \minor{\textit{removed\textsubscript{fix}}, \textit{removed\textsubscript{non-fix}},} or \textit{newly-introduced} between two consecutive revisions. Eventually, we crafted a dataset of \textbf{3,451} static code warnings and their evolution status for both manual analysis and future evaluation.     

After analyzing the limitations of the SOTA solution (\textbf{RQ2}), we propose \toolS to address the uncovered limitations. \toolS leverages refactoring information and the Hungarian algorithm~\citep{kuhn1955hungarian} to significantly improve the tracking precision. 
\minor{More importantly, we designed a heuristic-based algorithm in \toolS that can effectively decide whether a removed static warning is due to fix (\textit{removed\textsubscript{fix}}) or non-fix (\textit{removed\textsubscript{non-fix}}). The SOTA solution detects removed warnings and does not further categorize them as fix or non-fix.} 

\revised{ To answer \textbf{RQ3}, we first (RQ3.1) performed a comparative evaluation between StaticTracker and the SOTA approach.} Our evaluation of the collected 3,451 warnings shows that StaticTracker provides a significant improvement over the SOTA solution, i.e., \minor{the tracking precision improves from 64.4\% to 90.3\% for the tracking precision.} \revised{In RQ3.2, we conducted an independent evaluation of the performance of \toolS on a set of 2,014 commits.} 
\minor{The evaluation of RQ3.2 shows that \toolS achieves a precision of 90.2\% 
 in categorizing \textit{removed\textsubscript{fix}}, \textit{removed\textsubscript{non-fix}}, and \textit{newly-introduced} warnings. Particularly, it achieves a precision of 69.9\% in identifying removed warnings that are fixed by developers.}  

In summary, this paper makes the following contributions:
\begin{itemize}
	\item We collected and manually labeled a dataset of 3,451 static code warnings and uncovered the ground-truth evolution status between two consecutive commits. The static code warnings are detected by two mature and widely used static bug detectors (\textit{PMD} and \textit{Spotbugs}) on four real-world open-source software projects (\textit{JClouds}, \textit{Kafka}, \textit{Guava}, and \textit{Spring-boot}).
	\item We examined the SOTA solution in tracking the evolution of static code warnings in terms of tracking accuracy based on the collected dataset. Our investigation shows that the SOTA solution achieves inadequate results.
	\item We performed a manual analysis to uncover the inaccuracies and the reasons behind the low accuracy of the SOTA solution. Our findings offer empirical evidence to further improve the tracking of static code warnings in the development history.
	\item We proposed a better approach \toolS to tracking static code warnings in development history. \minor{In addition to a much higher tracking precision, \toolS includes a heuristic-based algorithm to 
 further categorize fixed and non-fixed warnings among the removed warnings.} The evaluation based on the crafted dataset shows that StaticTracker can significantly improve tracking precision.
\end{itemize}

\noindent\textbf{Artifact.} 
\junjie{We provide a replication package \footnote{\url{https://doi.org/10.5281/zenodo.6549386}}. Our replication package includes the implementations of the SOTA approach and StaticTracker, as well as our manually labeled ground-truth data for future evaluations. }

\noindent\textbf{Paper organization.} The rest of the paper is organized as follows. Section~\ref{sec:background} describes the motivation and the background, i.e., the relevant knowledge on static code warnings and how the SOTA approach works to track the static code warnings in development history. Section~\ref{sec:manual} illustrates the process and results of our manual analysis to understand the problems of the SOTA solution, including how the dataset is crafted and what are the insufficiencies of the SOTA solution. Section~\ref{sec:method} shows the details of \textbf{StaticTracker}, and its evaluation is shown in ~\ref{sec:evaluation}. Section~\ref{sec:threats} describes the threats to validity. Section~\ref{sec:related} lists the related work, and Section~\ref{sec:conclusions} concludes this study and proposes future works.

\section{Motivating Examples and Background}
\label{sec:background}
\revised{In this section, we formulate the problem we aim to solve, which is to provide a better approach to tracking the static code warnings in development history. Also, we provide background knowledge on the basics of static code warnings and how the state-of-the-art (SOTA) solution proposed by Avgustinov et al.~\cite{trackingSpacco} is used for tracking the evolution of static code warnings.}
\begin{figure}[h]
  \centering
\scalebox{0.95}{
  \includegraphics[width=\linewidth]{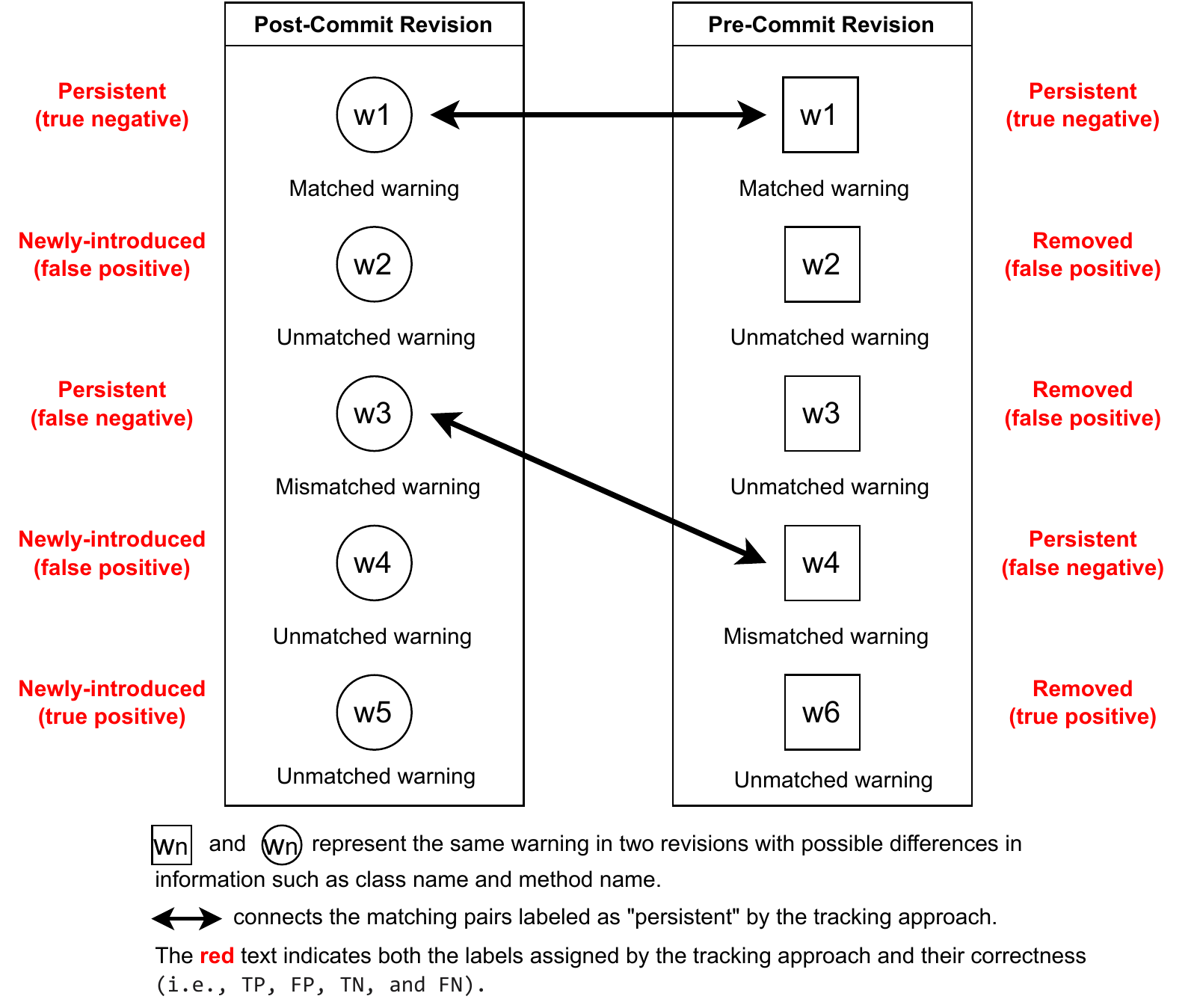}
}
  \caption{\minor{An example to show how the SOTA approach may produce false positives and false negatives due to incorrect mappings. Note that the SOTA approach only reports the combined status \textit{removed} rather than \textit{removed\textsubscript{fix}} and \textit{removed\textsubscript{non-fix}} separately.}}
  \label{fig:matchingWarning}
\end{figure}



\subsection{Challenges of tracking static code warnings in development history}
\label{sec:tracking_background}
Static bug detectors report a list of static code warnings given one version of a software system (i.e., one revision).  
\revised{A static code warning is subject to change as code evolves.
Similar to tracking code changes,} tracking the evolution of static code warnings in development history is based on comparing the generated reports from every two consecutive revisions. \revised{Applying a tracking solution, such as the SOTA approach, finds mappings of static code warnings between every two executive revisions and categorizes each static code warning to one of the \minor{four} following statuses, i.e., \minor{\textbf{removed\textsubscript{fix}}, \textbf{removed\textsubscript{non-fix}}, \textbf{newly-introduced}, and \textbf{persistent}. Note that in the rest of this paper, we use \textbf{removed} to denote the combination of \textbf{removed\textsubscript{fix}} and \textbf{removed\textsubscript{non-fix}}}}. 
\begin{itemize}
            \item \minor{\textbf{Removed\textsubscript{fix}}: A warning from the pre-commit revision is removed in the post-commit revision due to being fixed by developers.}
            \item \minor{\textbf{Removed\textsubscript{non-fix}}: A warning in the pre-commit revision is removed in the post-commit revision due to a non-fix reason, such as code deletion.}
         
	\item \textbf{Newly-introduced}: A warning is not in the pre-commit revision, and later is introduced in the post-commit revision.
	\item \textbf{Persistent}: A warning does not change between the pre-commit and post-commit revision.
\end{itemize}

\noindent\textbf{A good tracking mechanism should precisely label each static code warning as either persistent, \minor{removed\textsubscript{non-fix}, removed\textsubscript{fix}} or newly-introduced.}
\revised{In particular, it requires that all the mappings of static code warnings across two executive revisions are correctly established despite that the information of static code warnings used for mapping may be modified as code evolves. For example, the file name and code line number of the same warning may differ in consecutive revisions, which is challenging to find a correct mapping. }
Various types of software maintenance efforts contribute to making the tracking problem more complicated than one may imagine.
For example, \revised{code changes that are irrelevant to efforts of resolving static code warnings, such as a drastic refactoring, may modify the metadata information (such as start line, end line, or class name) of static code warnings, which makes the tracking tool map warnings incorrectly due to the different metadata of a warning between two revision.}

\revised{We use a simplified example (Figure~\ref{fig:matchingWarning}) to explain the challenges of tracking the static warnings between two consecutive revisions. 
\minor{For easy understanding, we use three statuses only, i.e., persistent, removed (including removed\textsubscript{fix} and removed\textsubscript{non-fix}), and newly-introduced.}
Given one commit, the left block represents all the warnings from the post-commit revision, and the right one is the pre-commit revision. Between the two revisions, the correct labels of the static warnings are as follows: persistent (\texttt{w1}, \texttt{w2}, \texttt{w3} and \texttt{w4}), \minor{removed} (\texttt{w6}), and newly-introduced (\texttt{w5}). 
Applying a tracking approach (i.e., the SOTA approach) establishes the mappings (shown as double arrow lines in Figure~\ref{fig:matchingWarning}). However, the established mappings are erroneous due to the limitations of the tracking approach, and result incorrect labels.
}

\revised{
For example, \texttt{w2} warnings from the two revisions are not correctly mapped by the SOTA approach due to code changes. 
As a result, \texttt{w2} is labeled as newly-introduced in the post-commit revision, while the correct label should be persistent.
Additionally, \texttt{w3} of the post-commit revision is incorrectly mapped with \texttt{w4} of the pre-commit revision. After the mappings are established, these warnings are classified into three statuses. \texttt{w2}, \texttt{w3} and \texttt{w6} in the pre-commit revision are labeled as \minor{\textit{removed}}. However, only one (i.e., \texttt{w6}) is correctly labeled among them, so it is a true positive. The others (\texttt{w2} and \texttt{w3}) are false positives of the tracking solution. Similarly, \texttt{w2}, \texttt{w4} and \texttt{w5} in the post-commit revision are labeled as \textit{newly-introduced}, but only \texttt{w5} is a true positive. \texttt{w2} and \texttt{w4} are false positives. The matched pairs are labeled as \textit{persistent} by the tracking approach. Both \texttt{w1} warnings in the post-commit and pre-commit revision are true negatives since they are mapped correctly. However, the labels of \texttt{w3} of the post-commit revision and \texttt{w4} of the pre-commit revision are matched incorrectly, so they are false negatives.
}

\minor{\pa{Evaluation Metrics.} We define a metric, namely precision, to evaluate the effectiveness of a tracking approach. 
For each status, we identify true positives (TP\textsubscript{[status]}), false positives (FP\textsubscript{[status]}), true negatives (FN\textsubscript{[status]}), and false negatives (FN\textsubscript{[status]}).
The precision is then calculated as $TP\textsubscript{[status]}/(TP\textsubscript{[status]} + FP\textsubscript{[status]})$. Below we give examples of these definitions using the status removed\textsubscript{fix}, fix for short.}

\begin{itemize}
	\item \minor{\textbf{TP\textsubscript{fix}}: A removed warning is identified as a fix by the tracking approach correctly.}
	\item \minor{\textbf{FP\textsubscript{fix}}: A removed warning is identified as a fix by the tracking approach incorrectly.}
	\item \minor{\textbf{TN\textsubscript{fix}}: A removed warning is identified as a non-fix by the tracking approach correctly.}
	\item \minor{\textbf{FN\textsubscript{fix}}: A removed warning is identified as a non-fix by the tracking approach incorrectly.}
\end{itemize}

\subsection{The Metadata of Static Code Warnings} 
Static bug detectors often represent detected static code warnings using metadata to help developers locate static warnings. Previous studies~\citep{Tracking15}~\citep{Fixpattern17} utilize the metadata information that static bug detectors provide for each warning to track the evolution of the detected static code warnings.   


Figure~\ref{fig:warningInstance} provides an example of a static code warning in \textit{JClouds} that is detected by \textit{Spotbugs}. We show the example of metadata in XML format. The metadata of the static code warning includes the following detailed information: the type of the static code warning (i.e., \textit{SE\_BAD\_FIELD}), and the problematic code region of this warning, which is represented by project name, class name, method name, field name, and the code range that is defined by a start line and an end line.
\begin{figure}[h]
		\lstset{
		language=XML,
		morekeywords={encoding,
			WarningType,WarningInstance,Commit,Class,Method,Field,Project,FilePath,StartLine,EndLine,WarningInstance, WarningType}
	}
	\begin{lstlisting}
	<WarningInstance> 
		<WarningType>SE_BAD_FIELD</WarningType>
		<Project>jclouds</Project> 
		<Class>ContextBuilderTest</Class>
		<Method></Method>
		<Field></Field>
		<FilePath>org/jclouds/ContextBuilder.java</FilePath>
		<StartLine>70</StartLine>
		<EndLine>75</EndLine> 
	</WarningInstance>
	\end{lstlisting}
	\caption{An example of the representation of one static code warning from Spotbugs. Note that the representation has been simplified to only show the information used by the SOTA matching approach.}
	\label{fig:warningInstance}
\end{figure}

\subsection{The state-of-the-art (SOTA) solution} 
Avgustinov et al.~\citep{Tracking15} proposed a multi-stage matching algorithm that can properly track the evolution of static code warnings under certain complicated software evolution, which we refer to as the SOTA solution. The overall structure of the SOTA solution is based on a pair-wise comparison between each warning in the pre-commit revision and each warning in the post-commit revision. Once a mapping is established, the two warnings from the two revisions are excluded for further comparisons. In particular, for each pair-wise matching process, i.e., between one warning from the pre-commit revision and one from the post-commit revision, four different matching strategies are placed in order, namely exact matching, location-based matching, snippet-based matching, and hash-based matching. 

\begin{algorithm}[h]
	\caption{\revised{The basic algorithm of the SOTA approach.}}
	\label{alg:SOA_Alg}
	\KwIn{The set of warnings from the pre-commit revision, $W_p$;
		The set of warnings from the post-commit revision, $W_c$;}
	\KwOut{The set of \minor{removed} warings, \minor{$W_{removed}$};
		The set of newly-introduced warings, $W_{newly-introduced}$;
		The set of matched pairs, $MatchedPairs$;}

	\For{each $w_i$ in $W_p$}
	{
		\For{each $w_j$ in $W_c$}
		{
			\If{source file of $w_i$ is not a changed file}
			{
				take $ExactMatching(w_i,w_j)$\;
			}
			\Else
			{
				take $ExactMatching(w_i,w_j)$\;
				
				take $LocationMatching(w_i,w_j)$\;
				
				take $SnippetMatching(w_i,w_j)$\;
				
				take $HashMatching(w_i,w_j)$\;
				
				\If{$w_i$ is matched up by any one of the four matching strategies}
				{
					make a $MatchedPair(w_i,w_j)$\;
					remove $w_j$ from $W_c$ \;
				}
			
			}
		}
		
	}
	$W_{newly-introduced} = W_c - MatchedPairs$\;
	
\end{algorithm}



				
		
	

Algorithm~\ref{alg:SOA_Alg} illustrates how the SOTA solution works to establish mappings between the list of warnings of two consecutive revisions. Exact matching requires every piece of metadata information to be matched and therefore is the most strict matching strategy among the four. When exact matching fails, the SOTA solution will then utilize the less strict matching strategy, i.e., location-based matching, which employes the diff algorithm to tolerate certain line shifts. If the location-based matching fails, the SOTA solution will continue to use snippet-based matching. When a class file was renamed or moved, the above matching strategies cannot handle that. Thus, the SOTA solution will utilize hash-based matching.   

At the end, when all the possible mappings are established, the unmatched warnings in the pre-commit revision are determined as \minor{removed}, and the ones in the post-commit revision are considered as newly-introduced.

\noindent\textbf{Exact matching.} 
Exact matching establishes the mappings for the warnings that are totally unaffected by the commit. For the two warnings, it is required that they have exactly the same source location (i.e., defined by the start and end line of the warning), warning type, and code information (i.e., class name, method name, and variable name).

\noindent\textbf{Location-based matching.} A commit may modify the information of certain static code warnings. Therefore when the exact matching fails, the following matching strategy, location-based matching, is used to tolerate the impacts to some extent. Location-based matching utilizes the \textit{diff} algorithms~\citep{hunt1977diff} ~\citep{myers1986diff} to derive source position mappings for each modified file. When a (potential) matching pair of warnings is in the diff output, location-based matching compares the offset of the corresponding warnings in the mappings. This matching requires the same warning metadata of code information (i.e., class name, method name, and variable name), but does not require the same source location (i.e., the start and end line of the warning).
If the difference of offsets is equal to or lower than \textit{3} (i.e., a fixed threshold), the location-based matching will decide the two warnings as a matching pair.  

\begin{figure}[hpt]
	\centering
	\scalebox{0.9}{
		\includegraphics[width=\linewidth]{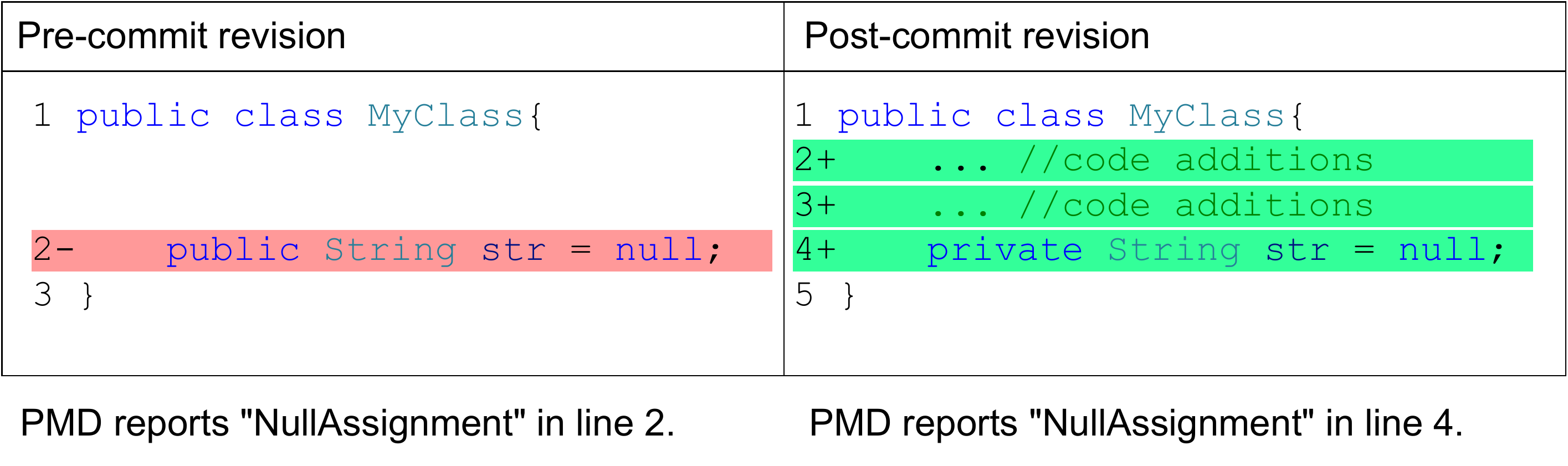}
	}
	\caption{\revised{An example to show how the location-based matching works to match the two ``NullAssignment" despite the different line numbers.}}
	\label{fig:exactMatching}
\end{figure}



As an example, Figure~\ref{fig:exactMatching} shows a diff mapping. The numbers on the left hand are the line numbers in the pre-commit revision. The numbers on the right are the line numbers in the post-commit revision. There is a PMD warning (`NullAssignment') reported in the pre-commit revision (line 2) and line 4 in the post-commit revision. Due to code adding, the source location (i.e., part of the warning metadata) has been changed. Location-based matching firstly computes the offsets between the source location and the diff mappings, respectively. The offset between the first line of the diff mapping (line 1) and the warning (line 2) is 1 for the pre-commit revision and 3 (line 1 and line 4) for the post-commit revision. Then, the differences between the two offsets are calculated. In this example, the difference is smaller than 3, so location-based matching will match the two warnings. 

\noindent\textbf{Snippet-based matching.} 
When code location changes significantly, the location-based matching approach may fail to identify persistent warnings across revisions. 
Snippet-based matching is used to address this problem. Given the source location defined by a start line and an end line, code snippets in between are extracted from both revisions. By performing the string matching on the two code snippets, snippet-based matching will decide a mapping if they are identical. Same as location-based matching, snippet-based matching requires the same warning metadata of code information (i.e., class name, method name, and variable name). \junjie{In comparison to location-based matching, snippet-based matching relies purely on code snippets, resulting in more precise matches. However, it has the disadvantage of being unable to match the warning when its code snippet has changed between revisions. For example, in Figure~\ref{fig:exactMatching}, there is a change from `public' to `private' on line 2 of the pre-commit revision and line 4 of the post-commit revision. Snippet-based matching will be failed to work in this case. In contrast, location-based matching can handle this scenario by disregarding the code snippet and attempting to locate near warnings based on the diffs. As long as the offset between the two warnings does not exceed a specific threshold, location-based matching can successfully match them.}

\noindent\textbf{Hash-based matching.} 
It is possible that a file is moved to a new location or renamed (i.e., class and file path are modified). Snippet-based matching cannot handle such cases well since the class name are required to be identical to perform snippet-based matching.
For such cases, a hash-based matching approach can be helpful. This matching approach tries to match warnings based on the similarity of their surrounding code. It first splits the text of the warning location into several tokens. Two hash values are calculated $h(W_{topN})$ and $h(W_{latter})$. $W_{topN}$ is $\mathit{n}$ tokens from the first one. $W_{latter}$ is tokens from the $\mathit{n+1^\text{th}}$ token to the last token. $\mathit{n}$ is a fixed threshold. \junjie{If the hash values of the top N or latter tokens of two warnings are identical, i.e., $h(W^{post}_{topN})=h(W^{pre}_{topN})$ or $h(W^{post}_{latter})=h(W^{pre}_{latter})$, they will be considered as a matched pair by the hash-based matching.}

\noindent\junjie{\textbf{Limitations of the SOTA approach.}
As we mentioned in Section~\ref{sec:tracking_background}, the metadata for the same warning may change in software evolution, while the matching strategies of the SOTA approach utilize the metadata in the matching process, which produces false positives of the tracked warnings from the SOTA approach. In this study, we improve the tracking approach to minimize the impact of the metadata changes of warnings in software evolution by employing three improvements.   
}

\section{Examining the performance of the SOTA Approach}

\label{sec:manual}
In this section, we describe how we investigated the performance of the SOTA approach in terms of the tracking accuracy, and answer \textbf{RQ1} and \textbf{RQ2}. In particular, we first crafted a dataset of static code warnings and their evolution status (i.e., persistent, \minor{removed\textsubscript{fix}, removed\textsubscript{non-fix}}, or newly-introduced) between two consecutive revisions. To craft this dataset, we re-implemented the SOTA approach, applied it to the development history of the studied open-source systems, and performed a manual analysis to determine the evolution status of each sampled static code warning. Then we manually analyzed whether the SOTA approach correctly tracked each sampled static code warning and categorized the reasons behind the incorrect tracking.

\subsection{Studied Subjects}
\label{sec:studySubjects}
\noindent\textbf{Static bug detectors.} \revised{We include two static bug detectors, i.e., \textit{PMD} and \textit{Spotbugs}, both of which are widely used in prior studies and adopted in practice. A study~\citep{lenarduzzi2021critical} compared six static analysis tools. Among them, PMD and Findbugs achieved promising precision (52\% and 57\%, respectively). Thus we adopt PMD and Spotbugs in this paper. In particular, Spotbugs, a spiritual successor of the well-known Findbugs, can detect more than 400 bug patterns in Java programs through bytecode analysis. Differently, PMD supports multiple languages and is known to be easily integrated into the build process.}

\noindent\textbf{Analyzed open-source systems.} \junjie{Our study includes four Java open-source systems, \textit{JClouds}, \textit{Kafka}, \textit{Spring-boot}, and \textit{Guava}. They are four popular Java projects in different areas. \textit{Spring-boot} is a framework for web applications, \textit{Kafka} is a distributed system to handle streaming data, \textit{Guava} is a core Java library for Google, and \textit{JCloud} is a cloud toolkit for the Java platform. Projects with different areas may be collected different static warnings. The four projects are used to summarize the insufficiencies of the SOTA approach and provide reasons for the introduction of false positives.} Then, they are used to evaluate StaticTracker and compare StaticTracker with the SOTA approach. We applied the two static bug detectors to all the revisions in a specific development period of the four studied software systems. We started with the official releases of the two software systems when we started this study, i.e., 2.3.1-rc2 of \textit{JClouds}, 2.1.0 of \textit{Kafka}, v2.3.6 of \textit{Spring-boot}, and v20.0 of \textit{Guava}. We selected the last commit of the studied release as the end date and its previous one and a half years as the studied development history. We were not able to successfully compile some revisions of systems in the studied period and excluded them from further studies. Besides, we also excluded the revision that has multiple pre-commit revisions.   
Table~\ref{tab:studiedSystems} lists the statistics of the studied systems, including the lines of code (LOC), the number of analyzed commits, the official release that we used to decide the end date of the studied development history, and the number of aggregated warnings in all the analyzed commits.

\begin{table}
\caption{\label{tab:studiedSystems} The studied systems and development periods. The release marks the end date of our studied development period, and we include 18 months of development history before the specified release.}
\scalebox{0.9}{
\begin{tabular}{c|rrrrr}
\toprule
& KLOC & \# Commits & Release & \multicolumn{2}{c}{\# Average Warnings}\\\cline{5-6}
& & & & PMD & Spotbugs\\\hline
Kafka & 434 & 2,000 & 2.3.1-rc2 & 12,972 & 27,911\\
JClouds  &494 & 300 & 2.1.0 & 18,176 & 2,090\\
Spring-boot  & 2,695 & 400 & v2.3.6 & 5,931 & 6,918\\
Guava  &2,112 & 2,000 & v20.0 & 6,474 & 3,819\\
\bottomrule
\end{tabular}
}
\end{table}
\subsection{Crafting the Dataset with Manual Labeling}
Before we describe how we collect the dataset, i.e., a list of static code warnings and their evolution status, 
we would like to motivate a few key points that drive our design choice in crafting the dataset.
First, given a large number of accumulative warnings across the revisions, we have to set our priorities, i.e., the evolution status of which static code warnings can better showcase the performance of the tracking approach since we have limited manual resources to spare.
Second, it is not surprising that in reality, the majority of the warnings persist in the codebases~\citep{not_use_static}. Therefore we do not consider it particularly interesting to include a corresponding percentage of persistent warnings in the dataset. 
Third, considering the downstream software engineering research this study can benefit from, we set our priorities to focus more on the static warnings that are \minor{\textit{removed}} or \textit{newly-introduced}.   

\junjie{Fourth, we choose to label all the warnings per sampled commit for constructing a ground-truth dataset, instead of the alternative, which is to sample warnings per commit and to include a larger set of commits, due to the inherent challenge in the mapping problem: one incorrect mapping may impact others, if only including one, we may only observe part of the impact, i.e., ``the tip of the iceberg'' as illustrated in Figure~\ref{fig:matchingWarning}.}  
Last, we have a decent confidence in the performance of the SOTA approach from its design. For example, we find that the majority of the established mappings by the SOTA approach are by the \textit{exact mapping} (e.g., 3,137 out of the 3,163 by Spotbugs in \textit{JClouds}-09936b5). The exact matching is the most strict matching process and rarely produces wrong results.

Guided by these key points, we decided to craft the dataset based on the tracking results of the SOTA approach and set our priorities on \minor{removed} and newly-introduced static code warnings. We apply the static bug detectors on a total of 4,700 commits in the four projects. 
\revised{We re-implemented the SOTA approach based on the released artifact by Liu et al.~\citep{Fixpattern17}. Moreover, since Liu's work focuses exclusively on Spotbugs, we had to implement the SOTA approach for PMD. Then, we applied the SOTA approach to track the evolution of the static code warnings across all the analyzed commits.} 

We select a subset of static code warnings for manual labeling following the steps below:
\begin{enumerate}
\item For \textit{JClouds\_Spotbugs}, \textit{JClouds\_PMD}, \revised{\textit{Spring-boot\_Spotbugs}, and \textit{Spring-boot\_PMD}} we include all the static warnings labeled as \minor{removed} by the SOTA approach.
\item Since there are many (i.e., 2,038 and 1,359) \textit{\minor{removed}} static warnings in \textit{Kafka\_PMD} and \textit{Kafka\_Spotbugs}, we took a statistically significant (95\%$\pm$5\%) sample, i.e., 326 and 301 \textit{\minor{removed}} static code warnings for both of them. Using \textit{Kafka\_PMD} as an example, we pursued the sampling process by firstly getting an estimation of the sample size, i.e., 323 warnings, then starting to randomly select one commit from the 436 \textit{Kafka} commits with at least one \minor{removed}  warning, until we collected more than 323 warnings. In the end, we collected 326 \minor{removed}  warnings from 53 commits in \textit{Kafka\_PMD}.

\item In \textit{Kafka\_Spotbugs}, \textit{Kafka\_PMD}, \textit{JClouds\_Spotbugs} and \textit{JClouds\_PMD}, there exist a large number of \textit{newly-introduced} code warnings. Hence, we took a statistically significant sample (95\%$\pm$5\%) of warnings in each setting and followed a similar sampling process as Step 2, i.e., including all newly-introduced warnings in one sampled commit. In the end, we collected a total of 704 warnings (i.e., in 47 commits) labeled as `newly-introduced' by the SOTA approach.

    \item \revised{We took the same sample strategy on \textit{Guava\_Spotbugs} and \textit{Guava\_PMD}, a statistically significant (95\%$\pm$5\%) sample of \minor{removed}  warnings with all newly-introduced warnings of their commits, i.e., 41 commits with 296 \minor{removed} warnings and 188 newly-introduced warnings in \textit{Guava\_PMD}, and 44 commits with 289 \minor{removed} warnings and 204 newly-introduced warnings in \textit{Guava\_Spotbugs}.}

\end{enumerate}

\begin{table}
\caption{\label{tab:sample} \revised{A summary of how we collect the static code warnings based on the results of the SOTA approach.}}
\scalebox{0.9}{
\begin{tabular}{crrrr}
\toprule
& \multicolumn{2}{c}{SOTA: \minor{Removed}} &\multicolumn{2}{c}{SOTA: Newly-Introduced} \\\hline

& \# Commits & \# Warnings & \# Commits & \# Warnings\\
\textit{\bf PMD} & &&&\\\hline
JClouds & 57 & 279 &19 & 155 \\
Kafka & 53 & 326 & 14 & 255\\
Spring-boot & 59 & 218 & 59 & 189 \\
Guava & 41 & 296 & 41 & 188\\
\hline
\textit{\bf Spotbugs} & \\\hline
JClouds& 23 & 104 & 5 & 78\\
Kafka & 26 & 301 & 9 & 216\\
Spring-boot& 17 & 193 & 17 & 160\\
Guava & 44 & 289 & 44 & 204\\
\hline
{\bf Total} & 320 & 2,006 & 208 & 1,445\\
\bottomrule
\end{tabular}
}
\end{table}

Table~\ref{tab:sample} summarizes the breakdown of the static code warnings we collected following the above-mentioned steps. 
Note that in Table~\ref{tab:sample}, the evolution statuses such as \textit{\minor{removed}} and \textit{newly-introduced} are labeled by the SOTA approach, which might be incorrect. We performed a manual analysis to reveal the true evolution status of each warning. \minor{The SOTA approach cannot detect whether a warning is removed for fix or non-fix reasons. 
Since one of our goals is to identify the warnings that are fixed by developers, we further manually categorize to removed\textsubscript{fix} or removed\textsubscript{non-fix}}.
Table~\ref{tab:dataset} summarizes the ground-truth evolution status of the dataset. In total, our dataset contains \revised{3,451} static code warnings and their true evolution status in the development history of the four projects: \revised{34.0\%} are persistent, \minor{4\% are removed\textsubscript{fix}, 33.1\% are removed\textsubscript{non-fix}}, and \revised{28.9\%} are newly-introduced.
In particular, two of the researchers individually performed a manual analysis to uncover the \textit{ground-truth} evolution status of each selected static code warning. The manual analysis includes understanding the nature of each static code warning and code changes that may involve the code warnings. The two researchers discussed the labels to resolve any disagreements. In our experiments, most of the disagreements are caused by human errors and can be easily agreed on. We calculated Cohen's kappa to measure the inter-rater agreement, which is the almost perfect level (0.96) in our experiment.  

It is noticeable that there exists a non-trivial discrepancy between Table~\ref{tab:sample} and Table~\ref{tab:dataset} regarding the distribution of the evolution statuses. That is because the SOTA approach produces a non-negligible number of incorrect results. We present more details on the inaccuracies in Section~\ref{sec:investigationFPs}.

\begin{table}

\caption{\label{tab:dataset}\revised{A data set of 3,451 static code warnings with manually-labeled evolution statuses. }}
\scalebox{0.9}{
\begin{tabular}{crrrr}
\toprule
& \textit{Persistent} & \minor{\textit{Removed\textsubscript{fix}}} & \minor{\textit{Removed\textsubscript{non-fix}}} & \textit{Newly-Introduced} \\
\hline
\textit{\bf PMD}& & & & \\\hline 
JClouds & 232  & 23 & 77 & 102\\
Kafka & 164 & 26 & 158 & 233\\
Spring-boot & 159 & 9 & 127 & 112\\
Guava & 257 & 4 & 163 & 60\\
\hline
\textit{\bf Spotbugs} &&& \\\hline
JClouds  & 32 & 29 & 56 & 65\\
Kafka & 205 & 15 & 174 &  123\\
Spring-boot & 12 & 1 & 186 & 154 \\
Guava & 113 & 31 & 199 & 150\\
\hline
\textbf{Total} & 1,174 & 138 & 1,140 & 999\\ 
\bottomrule
\end{tabular}
}
\end{table}

\begin{table}
\centering
\caption{\label{tab:SOAPerformance} \minor{The performance of the SOTA approach. Note that `removed' stands for both removed\textsubscript{fix} and removed\textsubscript{non-fix} as the SOTA approach detects three statuses only: removed, persistent, and newly-introduced.}}
\scalebox{1}{
\begin{tabular}{crrrrrr}
\toprule
&   \multicolumn{3}{c}{\minor{SOTA: Removed}} &  \multicolumn{3}{c}{\minor{SOTA: Newly-Introduced}} \\
&  \minor{TP}  & \minor{FP} & \minor{Precision} 
  & \minor{TP} & \minor{FP}  & \minor{Precision}  \\ \hline
\minor{\textit{\bf PMD}} &&& &&&\\\hline
\minor{JClouds} & \minor{100} & \minor{179} & \minor{35.8\%} & \minor{102}  & \minor{53} & \minor{65.8\%}\\
\minor{Kafka}  & \minor{184} & \minor{142} & \minor{56.4\%} & \minor{233} & \minor{22} & \minor{91.4\%}\\
\minor{Spring-boot}  &  \minor{136} & \minor{82} & \minor{62.4\%} &  \minor{112} & \minor{77} & \minor{59.3\%}\\
\minor{Guava}  &  \minor{167} & \minor{129} & \minor{56.4\%} & \minor{60} & \minor{128} & \minor{31.9\%} \\
\hline
\minor{\textit{\bf Spotbugs}} &&& &&&\\\hline
\minor{JClouds} &  \minor{85} & \minor{19} & \minor{81.7\%} & \minor{65}  & \minor{13} & \minor{83.3\%}\\
\minor{Kafka}  & \minor{189} & \minor{112} & \minor{62.8\%} & \minor{123} & \minor{93} & \minor{56.9\%}\\
\minor{Spring-boot}  &  \minor{187} & \minor{6} & \minor{96.9\%} &  \minor{154} & \minor{6} & \minor{96.3\%}\\
\minor{Guava}  &  \minor{230} & \minor{59} & \minor{79.6\%} & \minor{150} & \minor{54} & \minor{73.5\%}\\
\hline

\minor{\textbf{Total}}   & \minor{1,278} & \minor{728} & \minor{63.7\%} & \minor{999} & \minor{446} & \minor{69.1\%}\\

\bottomrule
\end{tabular}
}
\end{table}

\subsection{Evaluating the Performance of the SOTA Approach}
\label{sec:evalperf}

\begin{table}
	\caption{\revised{Causes of false positives by the SOTA approach.}}
	\label{tab:sixCauses}
	\scalebox{1.0}{
		\begin{tabular}{l|c}
			\toprule
			Cause & Number \\\hline
			1. Code refactoring & 437 \\
			2. Code shifting & 366\\
			3. Volatile class/method/variable names & 86\\
			4. Drastic and non-refactoring code changes & 285\\
			\hline
			{\bf Total} & 1,174 \\
			\bottomrule
		\end{tabular}
	}
\end{table}

\noindent\textbf{RQ1. Is the SOTA approach good at tracking the evolution of static code warnings? } 

We manually investigated the sampled \revised{3,451} static code warnings. Table~\ref{tab:SOAPerformance} summarizes the performance of the SOTA approach on the crafted dataset. Among the \revised{2,006} warnings that are determined as \textit{\minor{removed} } by the SOTA approach, only \minor{1,278 (63.7\%)} are \textit{truly \minor{removed}}. Among the \revised{1,445} warnings that are determined as \textit{newly-introduced}, only \revised{999 (69.1\%)} are \textit{actually} newly-introduced. \minor{The false positives (FPs) of ``removed" and ``newly-introduced" in Table~\ref{tab:SOAPerformance} are the warnings with a ground-truth status of \textit{persistent}}.
In short, the \minor{precision in tracking both removed and newly-introduced warnings} of the SOTA approach on the collected dataset is only \minor{66.0\%} \minor{(2,277/3,451)}. Our evaluation of the SOTA approach reveals that tracking the evolution of static code warnings over the development period is not that straightforward.
The low precision of the SOTA approach will negatively impact many downstream software engineering tasks, such as mining fix patterns from software repositories or performing empirical studies on software quality.

To this end, we answer \textbf{RQ1} after examining the performance of the SOTA approach by analyzing a number of tracked static code warnings.   

\rqbox{  
	We present a dataset of \revised{3,451} static code warnings and their evolution statuses. The dataset is crafted with support from the SOTA approach. The \minor{tracking precision} of the SOTA approach on the dataset is only \revised{66.0\%}, and this tracking approach will impact many downstream software engineering tasks negatively.}

\subsection{Investigating the Inaccuracies of the SOTA Approach}
\label{sec:investigationFPs}
\noindent\textbf{RQ2. What are the limitations of the SOTA approach?}  




\begin{figure}[hpt]
	\centering
	\scalebox{0.95}{
		\includegraphics[width=\linewidth]{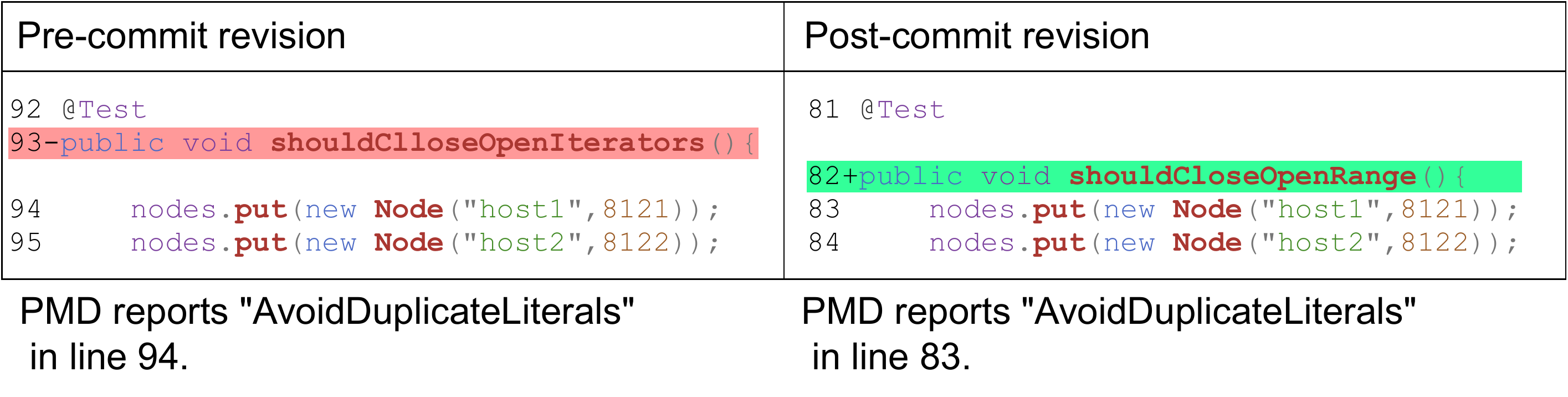}
	}
	\caption{\revised{An example of false positives due to method renaming.}}
	\label{fig:methodRenaming}
\end{figure}

 \junjie{Furthermore, we manually analyzed the insufficiencies of the SOTA approach, i.e., based on 1,174 false positives, and concluded into four categories as follows. Table~\ref{tab:sixCauses} summarizes four causes of false positives in our dataset.}
 
 \junjie{\noindent\textbf{Code refactoring.} We find that the SOTA approach cannot properly handle three common types of refactoring, namely class renaming, method renaming, and variable renaming. In particular, the first three matching strategies of the SOTA approach require the exact same class name, method name, and variable name, with a slight tolerance for line information. To tolerate minor differences in class/method/variable names (commonly caused by refactoring), the SOTA approach relies on the last matching strategy, namely the hash-based matching strategy. However, our experiment reveals that the hash-based matching strategy is sensitive to the regional code change, and fails to elegantly handle the refactoring changes since in reality, refactoring code changes are often combined with other code changes~\citep{RefactoringMiner18}. Note that we notice cases that are caused by more than one type of refactoring, e.g., one commit may contain both method renaming and variable renaming, which causes drastic changes in the metadata of warnings.}
 
 \junjie{Figure~\ref{fig:methodRenaming} is a case of a false positive due to method renaming. In this example, PMD detects a warning reported as `AvoidDuplicateLiterals' on line 94 of the pre-commit revision and line 83 of the post-commit revision. This warning indicates that the string `host1' is used multiple times in this class file. Due to the metadata of the method name changes, the first three matching strategies cannot match the warnings in the method. Thus the SOTA approach relies on the hash-based matching strategy. Due to the high sensitivity of the hash-based matching strategy, some persistent warnings in this method will not be mapped at all, which leads to inaccurate evolution statuses. Except for method renaming, other common code refactorings, such as class renaming and variable renaming also affects the consistency of the metadata between revisions. In total, we find 437 false positives in this category.}

\begin{figure}[hpt]
	\centering
	\scalebox{0.9}{
		\includegraphics[width=\linewidth]{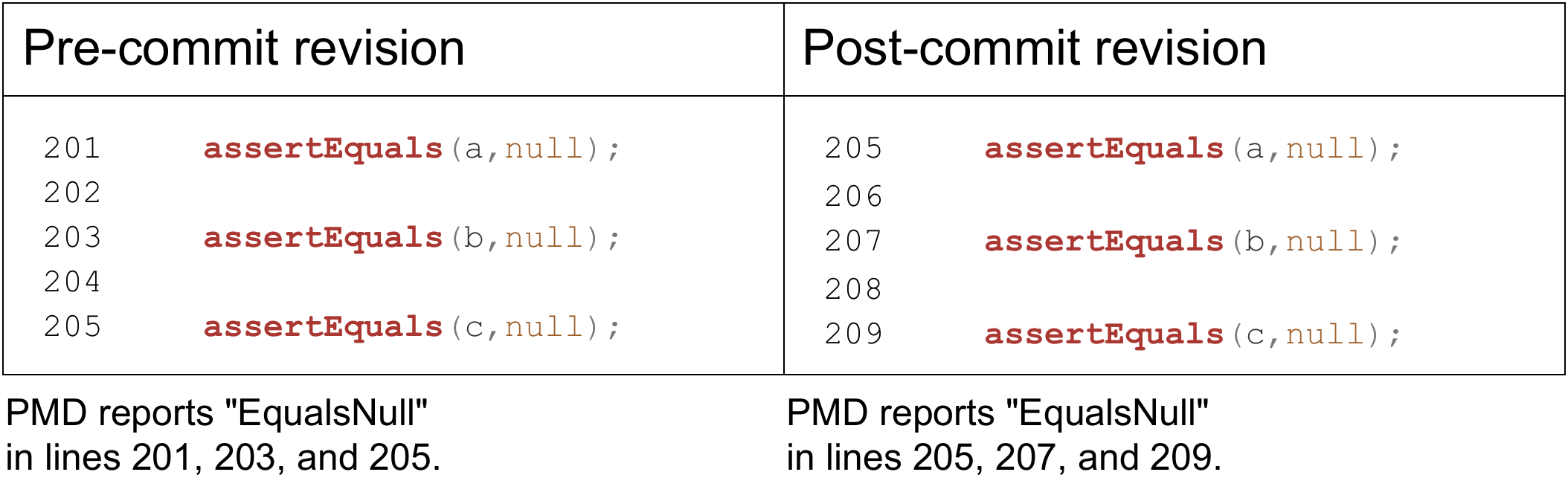}
	}
	\caption{\revised{An example of the false positives due to code shifting.}}
	\label{fig:mismatching}
\end{figure}




\noindent\textbf{Code shifting.}
Commits may modify the line numbers of some code statements, although these code statements are not directly modified by the commits. We call this code shifting. Because there exist similar code statements with similar static code warnings (e.g., same warning type, same variable, etc.), when code shifting happens, the SOTA approach does not always handle the shifting well, and false positives will be produced. Totally, we find 366 false positives in this category.

Figure~\ref{fig:mismatching} shows an example of how code shifting may cause the SOTA approach to malfunction. \junjie{This example contains three warnings of ``EqualsNull" in lines 201, 203, and 205 in the pre-commit revision and lines 205, 207, and 209 in the post-commit revision. The task is a 3x3 mapping problem. When the SOTA approach tries to process this 3x3 problem, it first matches line 205 in the pre-commit revision with line 205 in the post-commit revision through the exact matching strategy (i.e., identical line number), because the highest priority of all the four matching strategies provided by the SOTA approach. This incorrect mapping (line 205 v.s., line 205) has a butterfly impact on the remaining mapping. For example,  when the SOTA approach tries to find a mapping warning for line 201 in the pre-commit revision,  it fails to match with line 205 in the post-commit revision since the latter has been matched with line 205 in the pre-commit revision. As a result, the SOTA approach produces false positives as lines 201 and 203 are labeled as \minor{removed}.
}

Even though the three statements remain unchanged, their line numbers become different. In the pre-commit revision, the line numbers are 201, 203, and 205, while in the post-commit revision, the line numbers are 205, 207, and 209. As a result, the warning in line 205 from the pre-commit revision is mapped with the warning in line 205 from the post-commit revision by exact matching. This incorrect mapping causes the warnings on lines 201 and 203 from the pre-commit revision to be considered as resolved, and lines 207 and 209 from the post-commit revision to be considered as newly-introduced warnings, while they actually persist.


\begin{figure}[h]
\begin{lstlisting}[language=diff,numbers=none,escapechar=!,basicstyle=\ttfamily\footnotesize]
groups.map(_ -> getAcl(opts, 
           Set(Read))).toMap[ResourcePatternFilter, Set[Acl]]
\end{lstlisting}

\caption{An example of Scala code that has implicit code changes. The meta data of the relevant warning from the pre- and post-commit revisions are shown in Figure~\ref{fig:xmlDollarMarks1} and Figure~\ref{fig:xmlDollarMarks2}.}
\label{fig:noChange1}
\end{figure}

\begin{figure}[h]
	\lstset{
		language=XML,
		morekeywords={encoding,
			WarningType,WarningInstance,Commit,ClassMethod,Field,Project,FilePath,StartLine,EndLine}
	}
	\begin{lstlisting}[escapechar=!]
	<WarningInstance> 
		<WarningType>SE_BAD_FIELD</WarningType> 
		<Project>kafka</Project> 
		<Class>AclCommand</Class>
		<Method></Method>
!\Hilight!		<Field>opts$4</Field>
		<FilePath>kafka/admin/AclCommand.scala</FilePath>
!\Hilight!		<StartLine>206</StartLine>
!\Hilight!		<EndLine>206</EndLine> 
	</WarningInstance>
	\end{lstlisting}
	\caption{The warning information from pre-commit revision }
	\label{fig:xmlDollarMarks1}
\end{figure}


\begin{figure}[h]
	\lstset{
		language=XML,
		morekeywords={encoding,
			WarningType,WarningInstance,Commit,Class,Method,Field,Project,FilePath,StartLine,EndLine}
	}
	\begin{lstlisting} [escapechar=!]
	<WarningInstance> 
		<WarningType>SE_BAD_FIELD</WarningType> 
		<Project>kafka</Project> 
		<Class>AclCommand</Class>
		<Method></Method>
!\Hilight!      <Field>opts$1</Field>
		<FilePath>kafka/admin/AclCommand.scala</FilePath>
!\Hilight!		<StartLine>330</StartLine>
!\Hilight!		<EndLine>330</EndLine> 
	</WarningInstance>
	\end{lstlisting}
	\caption{The warning information from post-commit revision}
	\label{fig:xmlDollarMarks2}
\end{figure}

\noindent\textbf{Volatile class/method/variable names.} Even though there are no explicit code changes in one commit, on certain files, the warning reports by \textit{Spotbugs}, which uses bytecode analysis, are sensitive to compilation. Although everything else remains unchanged, \textit{persistent} warnings across revisions may have different line numbers or different class/method/variables names.
Such differences will cause all the matching strategies to malfunction. This happens frequently in Scala code when anonymous classes and methods are used heavily. Then some persistent warnings are not matched correctly. Totally, we find 86 false positives in this category.

Figure~\ref{fig:noChange1} is an example of false positives even though there are no explicit code changes. The line number of the code line with a warning changes from 206 to 330. 
We examined the metadata of this warning across two revisions (Figure~\ref{fig:xmlDollarMarks1} and Figure~\ref{fig:xmlDollarMarks2}) and found that not only the line numbers are different, the variable names are also different (the differences are highlighted using blue lines in Figure~\ref{fig:xmlDollarMarks1} and Figure~\ref{fig:xmlDollarMarks2}).  

\noindent\textbf{Drastic and non-refactoring code changes.}
\junjie{In cases where the code change is significant, all the matching strategies applied by the SOTA approach may fail to function adequately. Such a scenario may arise, for instance, if the offset of the diffs surpasses the threshold, causing location-based matching to be failed. Similarly, the modified code snippet of the warnings makes snippet-based matching fail. The introduction of significant changes also poses a challenge to hash-based matching, as differing hash values are calculated between the two revisions.}

\rqbox{   
	
	We perform further manual analysis on the FPs of the crafted dataset, and identify four main causes behind the inaccuracies of the SOTA approach in tracking the evolution of static code warnings.}

\section{\tool: A Better Approach to Track Static Warnings}  
\label{sec:method}
\begin{algorithm}
	\caption{The algorithm of \tool.}
	\label{alg:improvedAlg}
	\KwIn{The set of warnings from the pre-commit revision, $W_p$;
		The set of warnings from the post-commit revision, $W_c$;}
	\KwOut{$W_{\minor{removed\_fix}}$ is the set of \minor{removed\textsubscript{fix}} warnings; $W_{removed\_non\_fix}$ is \minor{the set of removed\textsubscript{non-fix} warnings;} $W_{newly-introduced}$ is the set of newly-introduced warnings; $MatchedPairs$ is the set of matched pairs.}

	Construct $W^{hash}_c$, a hash index of $W_c$
	
	Initialize a Two-dimensional array $HMatrix$  
	
	Remove all Identifiers in $W_p$ and $W_c$
	
	\For{ each $w_i$ in $W_p$}
	{
		
		\If{ source file of $w_i$ is not a changed file}
		{
			take $ExactMatching(w_i,W^{hash}_c[h(W_i)])$\;
		}
		\Else
		{
			$w^{'}_i = refactoring(w_i)$; \qquad \algorithmiccomment{if there is no refactoring in the location of $w_i$, $w^{'}_i$ = $w_i$}.\\
			
			\For{ each $w_j$ in $W_c$}
			{
				\Else
				{
					take $SnippetMatching(w^{'}_i,w_j)$\;
					take $LocationMatching(w^{'}_i,w^{'}_j)$\;
					\If{there is any candidate from both approaches}
					{
						$HMatrix[i][j] += 1$\;
					}
				}

			}
		}
		
	}
	$MatchedPairs = Hungarian(HMatrix)$\;
	$W_{\minor{removed}} = W_{p} - MatchedPairs$\;
        \minor{$W_{\minor{removed\_fix}}$, $W_{removed\_non\_fix}$= identifyFixNonfixRemoval($W_{removed}$)\;
        //This function is detailed in Algorithm~\ref{alg:fixAlg}}
	$W_{newly-introduced} = W_{c} - MatchedPairs$\;
        
\end{algorithm}

Guided by our manual analysis results, we propose to improve the SOTA approach by better handling refactoring changes and revising a few key steps to improve the accuracy of irrelevant code changes. In particular, StaticTracker (as illustrated in Algorithm~\ref{alg:improvedAlg}) reuses the three matching strategies of the SOTA approach (i.e., Exact matching, Location-based matching, and Snippet-based matching) and revises a few key steps to improve the inaccurate tracking. \minor{In addition, we develop an algorithm to further distinguish fixes and non-fixes from the removed warnings, i.e., line 17 in Algorithm~\ref{alg:improvedAlg}.}


\noindent\textbf{Improvement 1 - Including refactoring.} We included the refactoring information to improve the tracking of static warnings using \junjie{RefactoringMiner 2.0~\citep{Tsantalis:TSE:2020:RefactoringMiner2.0}. RefactoringMiner 2.0 is the state-of-the-art tool to detect refactoring for Java language. RefactoringMiner 2.0 is shown to outperform RefDiff~\citep{silva2020refdiff} and GumTreeDiff~\citep{falleri2014fine} with a precision of 99.6\% and a recall of 94\%.}
We first created a replica of $w_i$ (namely $w^{'}_i$), which is from the pre-commit revision, and then modified the metadata of $w^{'}_i$ with the information from RefactoringMiner. For instance, if RefactoringMiner reveals that the class in $w_j$ is a result of refactoring of ``move and rename class'', we modify the class name in $w^{'}_i$ with the one after the refactoring activity to keep the consistent metadata of the warning from two revisions. Two of the matching strategies (i.e., snippet matching in line 10 and location matching in line 13) are re-applied to decide two warnings (i.e., $w_i$, and $w_j$) whether they are candidates of a matched pair. In particular, Hash-based matching is designed to handle the case of the class files renamed or moved that are included in refactoring information. Thus we remove hash-based matching. As of now, we include \textbf{22} types of refactoring that cause the modified metadata of warnings. 

		

\begin{figure}[h]
	\centering
	\scalebox{0.5}{
		\includegraphics[width=\linewidth]{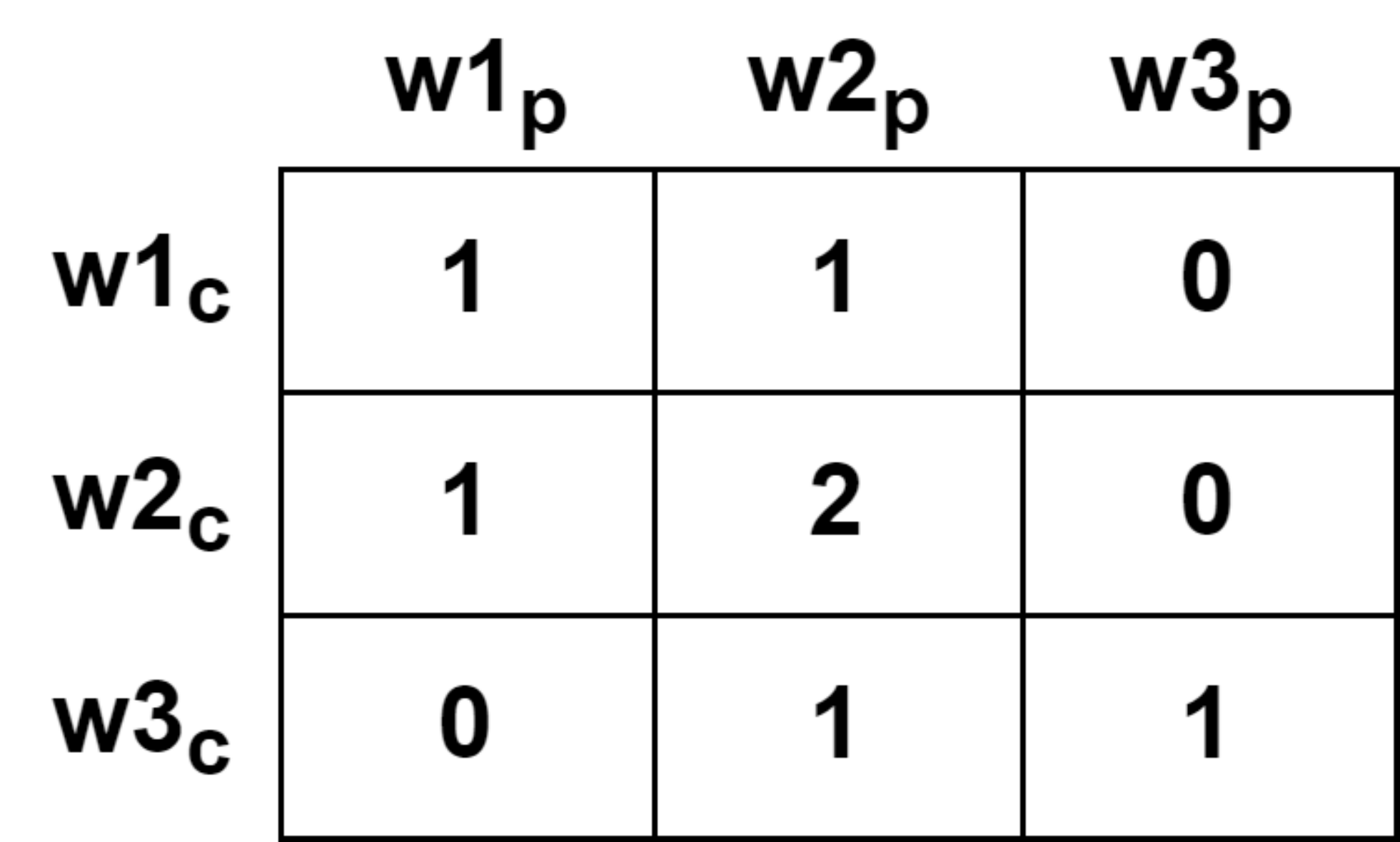}
	}
	\caption{A simple example of Hungarian matrix.}
	\label{fig:hungarianMatrix}
\end{figure}

\noindent\textbf{Improvement 2 - Deciding matched pairs using the Hungarian algorithm.} Commonly, a warning of pre-commit revision may have more than one matched warning from post-commit. Thus it is a problem of which one should be matched up. In the SOTA approach, it takes the first-come-first-matched, which may cause mismatching. Besides, the order of the matching strategies will affect the result. For example, we may get different results when we adopt location-matching first and snippet-matching first. The order in the SOTA approach is doing exact matching first, then location-based matching, and last one, snippet-based matching. In our investigation, this order has introduced many false positives like code shifting (Figure~\ref{fig:mismatching}). Besides, the first-matched warning may not be the best or correct one,i.e., there exist better-matched warnings. Thus we adopt \textbf{Hungarian algorithm}, a classic approach to solve the assignment problem in bipartite graphs. When a warning of post-commit revision is found that can be matched with a warning of pre-commit revision from the two matching strategies (i.e., location-based matching and snippet-based matching), instead of deciding it as a matched pair (i.e., a persistent warning), we construct a Hungarian matrix to save it as a potential matched pair. An example is like Figure~\ref{fig:hungarianMatrix}. $w1_p$, $w2_p$ and $w3_p$ are the warnings from pre-commit revision. $w1_c$, $w2_c$ and $w3_c$ are the warnings from post-commit revision. When two warnings are considered as a (potential) matched pair, the Hungarian matrix adds one (e.g., $w1_p$ and $w1_c$). A value of two (e.g., $w2_p$ and $w2_c$) means they are a (potential) matched pair from both matching strategies. It also means that this pair is more likely to be an actual pair of persistent warnings. If the SOTA is applied on the six static warnings , it is possible that $w1_p$ is matched with $w1_c$, and $w2_p$ is matched with $w3_c$, so $w3_p$ and $w2_c$ become false positives. In our algorithm, we construct a matrix $HMatrix$ (line 2) like Figure~\ref{fig:hungarianMatrix}. The size of $HMatrix$ is (the\ number\ of\ $W_p$) $*$ (the\ number\ of\ $W_c$) and the values are zero initially. Two matching strategies, snippet-based matching, and location-based matching are used to find out the potential matched warnings. Then we leverage maximum matching to decide the matched pairs.   
Besides, there is an exact matching for changed files in the SOTA matching, but if we adopt \textbf{Hungarian algorithm}, the matched warnings by exact matching can also be identified by location-based matching or snippet-based matching. Thus, we simply remove Exact matching for changed files in StaticTracker. However, we keep it for unchanged files.  

\noindent\textbf{Improvement 3 - Working with volatile identifiers.}
Anonymous classes and methods are given an identifier after compilation. However, the assigned identifiers are sensitive to change when there are code changes, even irrelevant. We try to minimize such sensitivity by removing the variable part in such identifiers. In particular, for identifiers such as \textit{opt\$1}, we use a regular expression to remove the numeric suffix after \$ and only keep \textit{opt} as the variable identifier in the metadata of a warning for the subsequent matching process.

\noindent\textbf{\minor{Improvement 4 - An approach to identify the removed warnings that are fixed by developers.}}
\minor{We further proposed a heuristic-based algorithm (described in Algorithm~\ref{alg:fixAlg}) that identifies fixes from removed warnings. \toolS applies GumTreeDiff~\citep{falleri2014fine} to extract the \textit{Diff} based on Abstract Syntax Tree (AST) representation. 
Algorithm~\ref{alg:fixAlg} takes a conservative way of identifying fixed warnings (line 8 and line 22) while identifying non-fix warnings proactively. 
For a given warning ($w_i$), the corresponding class, method, and field information are extracted (function \texttt{locate\_context} in line 2). If any of the class, method, and field declarations are completely deleted, then the warning is deemed \texttt{non-fix} (line 4). When $w_i$ is about the declaration of a source code entity (line 6),  we expect that a fix would be about modifying the declaration, such as removing \texttt{synchronized} from the modifiers. Alternatively, line 11--28 analyzes the detailed changes in the commit when $w_i$ is reported in a method ($mth$).
If $mth$ does not contain any code changes (line 13) or $mth$ only contains code deletions (line 15), $w_i$ is classified as non-fix.
When $w_i$ is about issues with a variable (line 19, a field name is reported in the metadata of $w_i$), our strategy is to identify whether there exists any modification on the reported field of $w_i$ (line 21). When $w_i$ is not about a particular field, our strategy is about trying to estimate a repair scope. If an estimated repair scope is not changed by the commit, i.e., no overlap between $Diffs$ and $mth$, our algorithm classifies this commit as non-fix proactively. By default, the repair scope is the method of the warning. For a few exceptional cases, our algorithm refines the repair scope further. In particular, such repair scope is estimated to include the range from the warning line ($w_i$.end) till the end of the method when $w_i$ is a one-line warning. 
Then, we check whether there are any modifications in the repair scope. If there are no modifications, $w_i$ is considered non-fix (line 28). Finally, the algorithm considers a fix for each of the remaining unlabeled warnings (line 30).
}
\newcommand{\colorizealgorithm}[1]{\color{black}#1\color{black}}
\begin{algorithm}
    \small
	\colorizealgorithm{\caption{\minor{\tool's algorithm to identify fix and non-fix warnings among removed warnings.}}}
	\label{alg:fixAlg}
 
	\colorizealgorithm{\KwIn{\minor{The set of removed warnings, $W_{removed}$; The commit diffs computed by GumTree, \texttt{Diff}}}}
	\colorizealgorithm{\KwOut{\minor{1. The set of removed\textsubscript{non-fix} warnings, $W_{non\_fix}$; \newline 2. The set of removed\textsubscript{fix} warnings, $W_{fix}$;}} }
        
        \colorizealgorithm{\ForEach{ $w_i$ in $W_{removed}$ }
        {
            \minor{(cls, mth, field) = locate\_context($w_i$);\\}
            \If{ is\_deleted(cls) 
            $\lVert$ 
            is\_deleted(mth)
            $\lVert$
            is\_deleted(field)
            } 
            {
                $W_{non\_fix}$.add($w_i$)\;
                \textbf{Next}\;
            }
            
            \If { same\_range($w_i$, cls) 
            $\lVert$
            same\_range($w_i$, mth)
            $\lVert$
            same\_range($w_i$, field)
            }
            {
             \If {is\_declaration\_modified(cls)
             $\lVert$
             is\_declaration\_modified(mth)
             $\lVert$
             is\_declaration\_modified(field)
             }
             {$W_{fix}$.add($w_i$)\;}
             \Else{ 
             $W_{non\_fix}$.add($w_i$)\;
             }
            }
            \ElseIf{ range($w_i$) $\in$ range(mth)}
            {
             repair\_scope = \{range(mth)\};\\
             diffs\_repair\_scope = \texttt{Diff} $\cap$ repair\_scope; \\ 
             \If{diffs\_repair\_scope == $\emptyset$} 
             {
                $W_{non\_fix}$.add($w_i$)\; 
             }
             \uElseIf{all\_deletions(\texttt{diffs\_mth})}
                {
                $W_{non\_fix}$.add($w_i$)\;
                }
            \Else{
           
              \If{field != \textbf{NULL} }
              { repair\_scope.append(range(mth));\\ 
              \If{(exists field is modified by Diffs $\cap$ repair\_scope)}
                {
                    $W_{fix}$.add($w_i$)\;
                } \Else{
                    $W_{non\_fix}$.add($w_i$)\;
                }
              } \uElseIf {$w_i$.start == $w_i$.end}
              {
                repair\_scope.append(\{($w_i$.end, mth.end)\});\\
                \If{\texttt{Diff} $\cap$ repair\_scope == \textbf{NULL}}
                {
                  $W_{non\_fix}$.add($w_i$)\;
                }
              }
            }
            }
        \If{$w_i$ $\notin$ $W_{fix}$ $\&\&$ $w_i$ $\notin$ $W_{non\_fix}$}
        {
            $W_{fix}$.add($w_i$)\;
        }
    }}
\end{algorithm}

\section{Evaluation of \tool}   
\label{sec:evaluation}
  
\noindent\textbf{RQ3. What is the performance of StaticTracker? }    
\junjie{We evaluated the performance of StaticTracker and set two sub-RQs for RQ3. The first one is a comparison evaluation between StaticTracker and the SOTA approach. We evaluated StaticTracker on the crafted dataset to show how much improvement StaticTracker has compared to the SOTA approach and answer \textbf{RQ3.1}. Furthermore, in \textbf{RQ3.2}, We re-sampled new commits and conducted an analysis of StaticTracker about how accurate it is on these commits.}

\subsection*{\revised{RQ3.1. Can StaticTracker perform better than the SOTA approach?}} 
\label{sec:comparative_analysis}

We compared the SOTA approach with StaticTracker by running both approaches on the same commits we labeled before. Since tracking the static code warnings is not a standalone task for each warning, it is, in fact, a mapping problem between two sets. Hence, we applied our improved approach to \textbf{all the warnings in the 320 commits}, which is a superset of the 3,451 warnings in the manually-labeled dataset. The remaining warnings in the 320 commits, while not in our crafted dataset, have a pre-assumed label, ``\textit{persistent}''. If StaticTracker changes the pre-assumed label of some warnings, then we manually examine the ground-truth labels of these warnings.

\begin{table*}[h]
\caption{\revised{Performance comparison between the SOTA approach and StaticTracker. \minor{Prec. is short for precision. A higher precision means fewer false positives and a better tracking performance.}}}
\label{tab:evaluation}
\centering
\scalebox{0.71}{
	
\begin{tabular}{crrrrrrrr}
\toprule
& \multicolumn{2}{c}{\minor{Removed\textsubscript{fix}}} & \multicolumn{2}{c}{\minor{Removed\textsubscript{non-fix}}} & \multicolumn{2}{c}{ Newly-Introduced} &
\multicolumn{2}{|c}{\bf Total} \\
 & &&& &&&  \multicolumn{2}{|c}{removed\textsubscript{fix}, removed\textsubscript{non-fix}, and newly-introduced}\\
\\\cline{2-9}
&  Prec. (SOTA) & Prec. (StaticTracker) & Prec. (SOTA) & Prec. (StaticTracker) & Prec. (SOTA) & Prec. (StaticTracker) & Prec. (SOTA) & Prec. (StaticTracker)\\
\textbf{PMD}  &&&& \\\hline
JClouds & 48.8\% (21/43) & 80.8\% (21/26) &              30.9\% (73/236) & 83.9\% (73/87) &                65.8\% (102/155) & 97.2\% (104/107) &                    45.2\% (196/434) & 90.0\% (198/220)\\ 
Kafka & 50.0\% (25/50) & 58.1\% (25/43) &              51.4\% (142/276) & 90.0\% (144/160) &                91.4\% (233/255) & 98.3\% (233/237) &                    68.8\% (400/581) & 91.4\% (402/440)\\\
Spring-boot & 56.2\% (9/16) & 69.2\% (9/13) &              60.9\% (123/202) & 85.4\% (123/144) &                59.3\% (112/189) & 86.7\% (111/128) &                    60.0\% (244/407) & 85.3\% (243/285)\\ 
Guava & 17.6\% (3/17) & 50.0\% (3/6) &              57.7\% (161/279) & 89.0\% (161/181) &                31.9\% (60/188) & 75.9\% (60/79) &                    46.3\% (224/484) & 84.2\% (224/266)\\\hline
\textbf{Spotbugs}  &&&& \\\hline
JClouds & 70.3\% (26/37) & 89.7\% (26/29) &              80.6\% (54/67) & 94.7\% (54/57) &                83.3\% (65/78) & 100.0\% (65/65) &                    79.7\% (145/182) & 96.0\% (145/151)\\
Kafka & 62.5\% (10/16) & 66.7\% (10/15) &              59.3\% (169/285) & 85.4\% (169/198) &                56.9\% (123/216) & 83.8\% (119/142) &                    58.4\% (302/517) & 83.9\% (298/355)\\\
Spring-boot & 50.0\% (1/2) & 100.0\% (1/1) &              97.4\% (186/191) & 100.0\% (186/186) &                96.2\% (154/160) & 100.0\% (154/154) &                    96.6\% (341/353) & 100.0\% (341/341)\\ 
Guava & 62.2\% (28/45) & 76.3\% (29/38) &              79.5\% (194/244) & 93.7\% (194/207) &                73.5\% (150/204) & 93.8\% (150/160) &                    75.5\% (372/493) & 92.1\% (373/405)\\\hline
\textbf{Total}  & 54.4\% (123/226) & \textbf{72.5\% (124/171)} &            61.9\% (1102/1,780) & \textbf{90.5\% (1,104/1,220)} &            69.1\% (999/1,445) & \textbf{92.9\% (996/1,072)} &                64.4\% (2,224/3,451) & \textbf{90.3\% (2,224/2,463)} \\
\bottomrule
\end{tabular}
}
\end{table*}

Table~\ref{tab:evaluation} shows the comparison results between the SOTA approach and our improved approach on the collected dataset of 3,451 static code warnings. Note that there are 3,451 static warnings from the SOTA approach. However, when we applied StaticTracker to the same dataset, we obtained only 2,463 \minor{removed} and newly-introduced warnings, which means that the rest are identified by StaticTracker as persistent warnings. We categorized the 3,451 warnings into \minor{three categories (i.e., removed\textsubscript{fix}, removed\textsubscript{non-fix}, and newly-introduced)} according to the labels by the SOTA approach for easy comparison. \minor{The SOTA approach does not detect if a removed warning is due to a fix. In order to ensure a fair and comprehensive comparison, we have integrated our fix-detecting algorithm (Algorithm~\ref{alg:fixAlg}) into the SOTA approach. The SOTA approach labels 226 fixes. Among them, 123 fixes are true positives. Therefore, the precision of the SOTA approach is 54.4\% for detecting fixes. The \toolS achieved a precision of 72.5\% (i.e., among 171 labeled fixes by \tool, 124 are true positives).} The evaluation shows that StaticTracker can significantly \minor{improve the tracking performance.} Overall, for the 3,451 warnings, the SOTA approach has \minor{1,227} warnings with wrong evolution statuses, i.e., \minor{the tracking precision is 64.4\%.} Compared to that, the precision of StaticTracker achieved 90.3\%. \textbf{StaticTracker reduces the false positives by correctly labeling the \textit{persistent} warnings, which are mistakenly labeled as \minor{\textit{removed}} or \textit{newly-introduced} by the SOTA approach.} 

StaticTracker is shown to effectively reduce false positives for \minor{the four causes listed from the SOTA approach in Table~\ref{tab:sixCauses}.} Table~\ref{tab:sixCausesAfter} shows the breakdown of the left false positives by each cause after using StaticTracker on the \minor{\textit{removed}} warning dataset. \minor{One more category is named `Fix-detection labels warnings incorrectly', which is caused by our proposed fix-detection approach.}

\begin{table}[h]
	\caption{\minor{A breakdown of \tool's false positives.}}
	\label{tab:sixCausesAfter}
		\begin{center}
	\scalebox{0.9}{
		\begin{tabular}{l|c}
			\toprule
			Cause & Number \\\hline
			1. Code refactoring & 78 \\
			2. Code shifting & 32\\
			3. Volatile class/method/variable names & 8\\
			4. Drastic and non-refactoring code changes & 68\\
                \minor{5. Fail to differentiate removed\textsubscript{fix} and removed\textsubscript{non-fix} warnings} & \minor{53} \\
			\hline
			{\bf Total} & \minor{239}\\
			\bottomrule
		\end{tabular}
	}
	\end{center}
\end{table}

\rqbox{
	\minor{\toolS outperforms the SOTA approach by reducing false positives significantly (i.e., from 1,227 to 239) and yields a precision of 90.3\% in detecting evolution statuses.}
 }

\subsection*{\revised{RQ3.2. How accurate is \textbf{StaticTracker} for tracking the evolution of static code warnings? }}   
Apart from the SOTA approach, we also take a generalization evaluation of StaticTracker by a statistically significant (95\%$\pm$5\%) sample on commits for each project to answer this RQ. StaticTracker is applied to collect \minor{removed} and newly-introduced warnings on sampled commits. Then two authors manually check them to determine whether a warning is a false positive or a true positive with Cohen's kappa coefficient of 0.82. \minor{Table~\ref{tab:performanceGeneralization} } shows the results of \toolS for RQ3.2. Note that there are many commits that have no \minor{removed} or newly-introduced warnings in this evaluation. In other words, the code changes of many commits are too small to change the status of all static warnings. We sampled 2,014 commits. \minor{Our approach correctly identified 51 removed\textsubscript{fix}, 339 removed\textsubscript{non-fix} and 794 newly-introduced warnings on these commits.} Overall, \toolS has a great performance \minor{in detecting evolution statuses with a tracking precision of 90.2\% (1,184/1,312). For detecting removed\textsubscript{fix} warnings, \toolS yields a precision of 69.9\%.} 

\begin{table}
	\caption{\label{tab:performanceGeneralization} \minor{The performance of StaticTracker (RQ3.2).}}
 
	\scalebox{0.7}{
		\begin{tabular}{crrrr}
			\toprule 
			& \# Commits & \minor{Prec. (Removed\textsubscript{fix})} & \minor{Prec. (Removed\textsubscript{non-fix})} & Prec. (Newly-Intr.) \\\hline
			\textit{\bf PMD} & &&\\\hline
			JClouds & 169 & 82.4\% (14/17) &              82.4\% (61/74) &                96.0\% (144/150)   \\
			Kafka & 322 & 26.7\% (4/15) &              91.0\% (101/111) &                94.8\% (145/153)  \\
			Spring-boot & 194 & 100.0\% (2/2) &              100.0\% (4/4) &                100.0\% (17/17)  \\
                Guava & 322 & 50.0\% (1/2) &                68.4\% (26/38) &                    91.7\% (100/109) \\
                \hline
			\textit{\bf Spotbugs} &&& \\\hline
			JClouds & 169 & 90.9\% (20/22) &              83.3\% (15/18) &                100.0\% (106/106)  \\
			Kafka & 322 & 50.0\% (4/8) &              75.8\% (69/91) &                90.3\% (168/186) \\
			Spring-boot & 194 & NA. (0/0) &                100.0\% (20/20) &                    100.0\% (10/10) ) \\
                Guava & 322 & 85.7\% (6/7) &                93.5\% (43/46) &                    98.1\% (104/106) \\
			\hline
			{\bf Total/Avg.} & 2,014 & 69.9\% (51/73)&            84.3\% (339/402) &             94.9\% (794/837)\\
			\bottomrule
		\end{tabular}
	}
\end{table}


\rqbox{
	By conducting the generalization analysis of \tool, results show that \toolS achieves a \minor{tracking precision of 90.2\% in identifying evolution status of warnings.} 
}

\subsection*{\junjie{Discussions}}
\label{sec:generalization_analysis}
\junjie{We provide further discussions on 1) the false positives of our proposed approach \toolS and the reasons behind such false positives; 2) an ablation analysis on the three improvements we designed for \tool; and 3) correlations between the warning types and the number of warnings tracked correctly or not.} 

\noindent\textbf{\junjie{Analysis on the false positives of \tool.}}
\junjie{
We conducted a detailed investigation into the false positives in StaticTracker and discussed them. To achieve this, we sampled \minor{75} warnings from the total of \minor{239} warnings with a statistically significant
(95\%$\pm$10\%) sample and manually analyze each one to uncover its root causes. We summarize the uncovered causes below. 
}

\begin{figure}[hpt]
	\centering
	\scalebox{0.95}{
		\includegraphics[width=\linewidth]{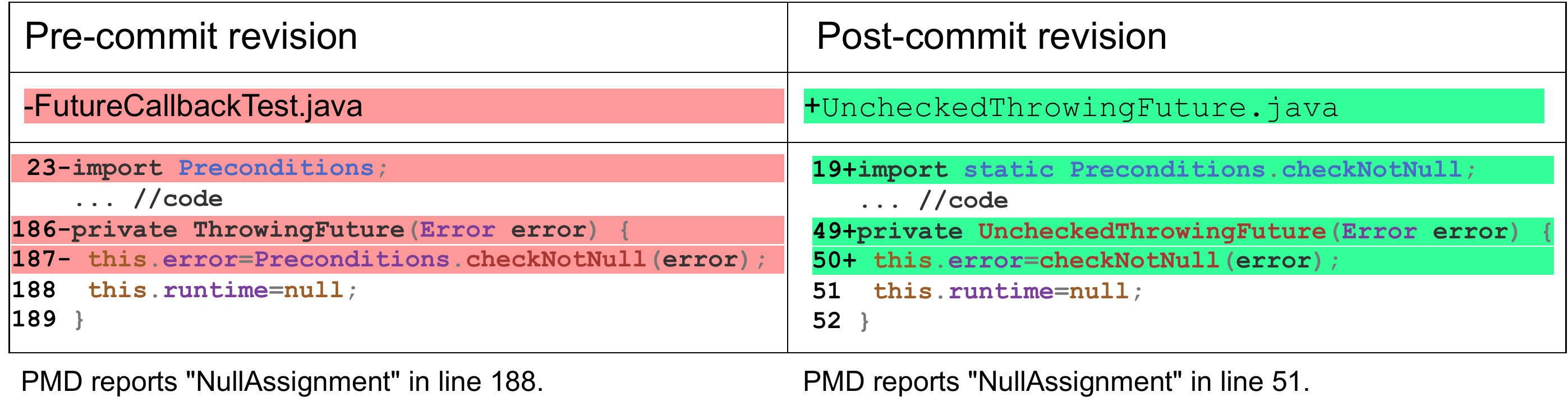}
	}
	\caption{\revised{An example of the false positive in StaticTracker due to undetected method renaming by RefactoringMiner. The commit is ba2024d from Guava.}}
	\label{fig:fp_our_refactoring}
\end{figure}

\minor{- \underline{\it Undetected refactoring (32/75)}.} \junjie{Even with the state-of-the-art refactoring detection tool (i.e., RefactoringMiner), there exist refactoring changes that are not detected. This contributes to almost half of the false positives of \tool. Figure~\ref{fig:fp_our_refactoring} shows an example of unmatched static warnings due to undetected refactoring, i.e., the same warning of `NullAssignment' (line 187 in the pre-commit revision and line 51 in the post-commit revision).   
The commit includes one class renaming and one method renaming and the latter (from \texttt{`ThrowingFuture'} to \texttt{`UncheckedThrowingFuture'}) is not detected by RefactoringMiner. 
As a result, \toolS fails to match the same warning from the two consecutive revisions and labels the warning incorrectly as \minor{removed} and newly-introduced, respectively.}

\begin{figure}[hpt]
	\centering
	\scalebox{0.95}{
		\includegraphics[width=\linewidth]{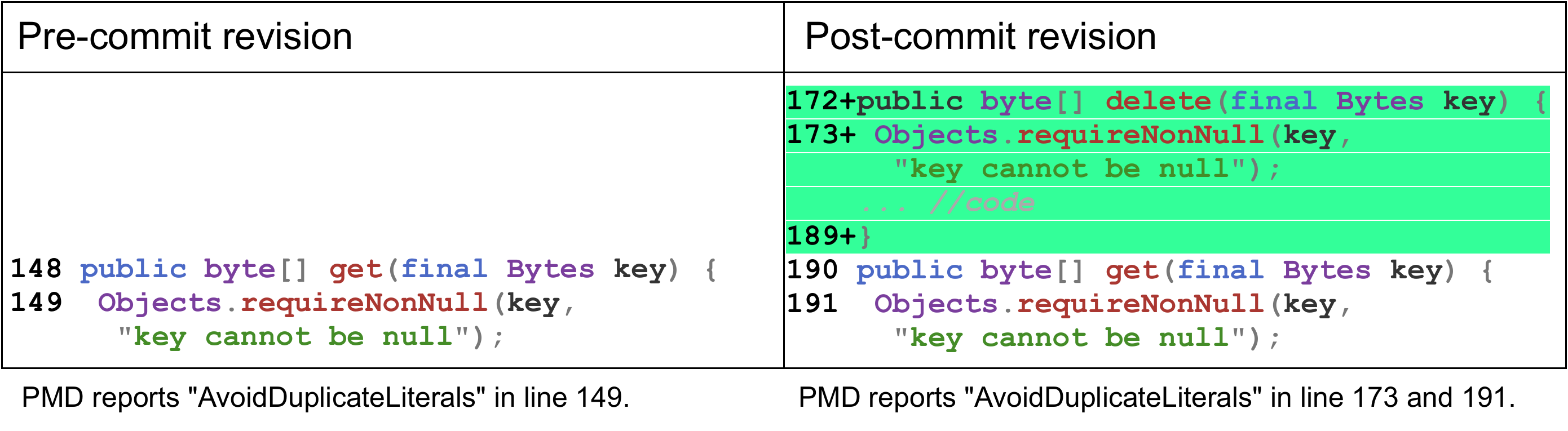}
	}
	\caption{\junjie{An example of a warning in the pre-commit revision superseded by a new warning in the post-commit revision. The commit is 3c46b56 from Kafka.}}
	\label{fig:fp_our_moved_warning}
\end{figure}

\minor{- \underline{\it Superseded by a new warning (15/75)}.} \junjie{We find for 15 cases, the warnings in the pre- and post-commit revisions are highly similar, i.e., one is superseded by the other as code evolves. Figure~\ref{fig:fp_our_moved_warning} shows an example of a warning `AvoidDuplicateLiterals' detected by PMD. The string ``key cannot be null" is used multiple times and PMD reports one warning for the multiple uses of the duplicate literal. This warning is then superseded by a highly similar warning when the new code contains another use of the duplicate literal. The warning in the post-commit revision contains a different location, i.e., from line 191 in the method \textit{get} to line 173 in the method \textit{delete}. As a result, \toolS fails to match the two correctly. 
Another type of example is about one warning is superseded by another warning of a different but related warning type. Detecting the two warning types share some similarities. Some code changes irrelevant to the scope of the reported warning may easily alter the detection from one type to another type. For example, after Guava-5562218, one warning of `SE\_BAD\_FIELD\_INNER\_CLASS' is changed to `SE\_INNER\_CLASS' due to a code change to the outer class, which is not on the reported code scope (i.e., the inner class). 
}

\minor{- \underline{\it Limitations of using Hungarian algorithm (4/75)}.} 
\junjie{Usually, the Hungarian algorithm is effective in establishing matching pairs. However, certain situations may arise wherein a warning from the pre-commit revision has two possible candidates from the post-commit revision, and the two possible candidates have equivalent weights. For such cases, the Hungarian algorithm may fail to accurately determine the matched pair, resulting in a mismatch.}


\minor{- \underline{\it Drastic code changes (16/75)}.}
\junjie{Metadata of static warnings used for matching is changed significantly by drastic code changes. The two matching strategies (i.e., location and snippet matching) are designed to effectively match most persistent warnings by tolerating non-drastic code changes. When there is a significant change in the class file, both strategies may fail to match certain warnings, resulting in false positives. }  

\minor{- \underline{\it Limitations of the fix-detection approach (8/75)}.}
\minor{Our proposed fix-detection approach cannot detect every fix and non-fix case correctly. Among the 75 false positives, eight cases are identified with a wrong status between removed\textsubscript{fix} and removed\textsubscript{non-fix}. 
}

\noindent\textbf{\junjie{Ablation analysis.}}
\junjie{We proposed three improvements in StaticTracker to tackle the limitations of the SOTA approach: 1) Handling volatile identifiers (VI), 2) Detecting refactoring using RefactoringMiner (RM), and 3) Matching using the Hungarian algorithm (HA).}

\junjie{To evaluate the impact of the improvements (individual and combined), we performed an ablation analysis. Since VI ``correct'' the metadata of static code warnings and RM and HA improve the matching process, we consider VI is fundamental and select combinations on top of VI: 1) VI, 2) VI+RM, and 3) VI+HA. Table~\ref{tab:decomposed_improvements} presents the false positive rates of these different combinations of the three improvements. Our findings demonstrate that handling volatile identifiers results in a slight \minor{ increase to the tracking precision, i.e., from 64.4\% to 67.9\%.} Additionally, both VI+RM and VI+HA approaches have similar performance, with \minor{precisions of 75.6\% and 76.6\%}, respectively. Notably, the combination of all three improvements in StaticTracker resulted in \minor{a significant improvement of the tracking process}, at \minor{90.3\% precision}. In short, the proposed three improvements complement each other, and the combination of all three significantly outperforms the other combinations.}

\begin{table}[h]
	\caption{\junjie{Ablation analysis on the three improvements \toolS has over the SOTA approach.}}
	\label{tab:decomposed_improvements}
		\begin{center}
	\scalebox{1}{
		\begin{tabular}{l|r}
			\toprule
			Approach & \minor{Precision}\\\hline
			Baseline (SOTA) & \minor{64.4\% (2,224/3,451)} \\
			Baseline + \textbf{VI} & \minor{67.9\% (2,211/3,256)}\\
			Baseline + \textbf{VI}+\textbf{RM} & \minor{75.6\% (2,210/2,923)}\\
			Baseline + \textbf{VI}+\textbf{HA} & \minor{76.6\% (2,220/2,900)}\\
			Baseline + \textbf{VI}+\textbf{RM}+\textbf{HA} (\tool) & \minor{90.3\% (2,224/2,463)}\\
			\bottomrule
		\end{tabular}
	}
	\end{center}
\end{table}

\noindent\textbf{\junjie{Correlation between the performance of \toolS and the types of static warnings.}}
\junjie{
 We analyzed whether there exist significant different performance improvements of \toolS over the SOTA approach on each warning type involved. Particularly, this evaluation involves 32 PMD warning types and 111 Spotbug warning types. We performed Fisher's exact test on the pair of true positives and false positives per warning type between the two approaches (\toolS and SOTA). We find that for seven PMD warning types and six Spotbugs warning types, the differences between \toolS and Spotbugs are statistically significant, i.e., the improvement of \toolS on these warning types is significant. Table~\ref{tab:correlation_warning_type} describes the detailed results.
}

\minor{\noindent\textbf{False Negatives.} The false negatives in our context are the warnings deemed as `persistent' by \toolS have a ground-truth label of either removed\textsubscript{fix}, removed\textsubscript{non-fix}, or newly-introduced. It is extremely time-consuming and very challenging to identify the false positives in our context due to the tremendous number of \textit{persistent} warnings (e.g., up to thousands of warnings between two consecutive versions).
We believe both the SOTA and \toolS have low numbers of false negatives since both have highly strict rules to decide persistent warnings, i.e., the metadata of two warnings have to be highly similar to be considered for matched warnings between two consecutive versions.
To provide evidence, we took a statistically significant sampling (95\%$\pm$5\%) of 384 persistent warnings from a total of over six million persistent warnings. Subsequently, we analyzed the sampled warnings and confirmed that all of them were true negatives, i.e., no false negatives are identified from this statistically significant sample. This demonstrates that \toolS likely has very high recall in identifying evolving warnings.}  
\begin{table}[h]
\caption{\junjie{Correlation between the performance of \toolS in comparison with the SOTA approach and warning types.}}
\label{tab:correlation_warning_type}
\centering
\scalebox{0.55}{
	
\begin{tabular}{lrrrrr}
\toprule
& \multicolumn{2}{c}{SOTA} 
& \multicolumn{2}{c}{StaticTracker}  
\\\hline
& \minor{TP} & \minor{FP} & \minor{TP} & \minor{FP} & p-value\\\hline
\textbf{PMD}  &&&&& \\\hline

BeanMembersShouldSerialize & 444 & 276 & 443 & 35 & 4.2e-37\\
DetachedTestCase & 5 & 17 & 7 & 0 & 5.0e-4\\
AvoidDuplicateLiterals & 121 & 220 & 121 & 50 &5.5e-14\\
DataflowAnomalyAnalysis & 301 & 164 & 300 & 41 &2.1e-14\\
AvoidFieldNameMatchingMethodName & 64 & 62 & 66 & 2 & 1.7e-12\\
NullAssignment & 16 & 17 & 17 & 4 & 0.02\\
AvoidCatchingNPE & 3 & 26 & 3 & 0 & 0.004\\

\hline

\textbf{Spotbugs}  &&&&& \\\hline
DLS\_DEAD\_LOCAL\_STORE & 54 & 21 & 53 & 4 & 0.003\\
NP\_PARAMETER\_MUST\_BE\_NONNULL\_BUT\_MARKED\_AS\_NULLABLE & 49 & 10 & 49 & 2 & 0.03 \\
SE\_BAD\_FIELD & 102 & 86 & 102 & 18 &1.6e-8\\
SIC\_INNER\_SHOULD\_BE\_STATIC\_ANON & 121 & 77 & 122 & 10 &3.1e-11\\
NP\_ALWAYS\_NULL & 38 & 60 & 37 & 8 &1.6e-6\\
NP\_LOAD\_OF\_KNOWN\_NULL\_VALUE & 26 & 32 & 25 & 8 &0.004\\
\bottomrule
\end{tabular}
}
\end{table}

\section{Threats to validity}
\label{sec:threats}
In this section, we describe threats to external and internal validity.
\subsection{External Validity}
In this paper, we focus on tracking the static code warnings in Java projects. Our study results may not be generalizable to projects in other languages. It is expected that programs with similar evolution details to Java systems may benefit from our study. We include two static bug detectors in our study, whose representation of static code warnings are similar to some extent, i.e., the use of class/method/variable names and code ranges for matching purposes. The improvement of StaticTracker may not be generalizable to a static bug detector with a different set of metadata of the reported warnings. However, most of the popular static bug detectors provide similar information.  

Last, our crafted dataset for evaluating and improving the SOTA approach is based on four open-source projects. To increase the diversity, we analyzed a reasonable number of commits in the four projects. In general, we find that the evolution details that make the SOTA approach malfunction are consistent in our collected dataset. In the generalization analysis, we sampled commits to evaluate StaticTracker, but many commits have no disappeared and newly-introduced warnings, which means that all warnings from these commits are labeled as persistent warnings by StaticTracker.

\subsection{Internal Validity}
When it comes to manually labeling the dataset, human errors are inevitable. We tried to reduce human errors by having two people annotating the dataset and resolving conflicts through discussions.

Although our dataset covers warnings with all three evolution statuses, we do not claim that our dataset is representative in terms of following the distribution of the three evolution statuses.   

In particular, we set our criteria in crafting the dataset based on our observations on the SOTA approach (i.e., most of the established mappings are correct) and also our priorities, which is to focus on the disappeared and newly-introduced warnings.

\section{Related Work}
\label{sec:related}
\noindent\textbf{Tracking the evolution of code issues.}
Tracking the evolution of code issues, whether bugs, code smells, or static code warnings, is a central question in many software quality studies. For example, the \textit{SZZ} algorithm~\citep{sliwerski2005changes}, which identifies the origin of bug-introducing commits, is widely used in defect prediction studies. Recent evaluations have uncovered many previously unknown deficiencies in \textit{SZZ} and inspired many researchers to work on improving \textit{SZZ}. For example, a study ~\citep{refactoring_szz} empirically investigated how bug-fix changes and bug-introducing changes of the \textit{SZZ} are impacted by code refactoring. Then they proposed refactoring-aware \textit{SZZ}. Another study~\citep{mz_szz} proposed a framework to provide a systematic evaluation of the data collected by \textit{SZZ}. Palix et al. conducted two studies on mining the code patterns. The first study~\citep{palix2010tracking} presented a language-independent tool for mining and tracking code patterns across the evolution of software by building graphs and computing statistics. Their other study~\citep{palix2015improving} combined the tool with AST for the detection of code patterns across multiple versions. There is a study~\citep{querel2018warningsguru} that presented a tool that combines static analysis with statistical bug models to detect which commits are likely to contain risky codes, which provides more precise information about a static warning.
Dong-Jae et al.~\citep{annotation} conducted an empirical study on the evolution of annotation changes and created a taxonomy to uncover what annotation changes have and the motivation of annotation changes. 
In addition, Felix et al.~\citep{grund2019codeshovel} proposed a tool to uncover method histories with no pre-processing or whole-program analysis, which quickly produces complete and accurate change histories for 90\% of methods.  

Compared to tracking the defects, tracking the static code warnings has been increasingly needed in recent research, yet rarely studied for its challenges and insufficiencies. 
Spacoo et al.~\citep{trackingSpacco} propose to match warnings across revisions using a combination of some basic information of each warning (e.g., warning type, class/method names) and allow inexact matching to some extent. Their approach is not able to match warnings if they are moved to a different class/method. Other diff-based approaches are used to identify which static code warnings are disappeared. In particular, Sunghun et al.~\citep{kim_tracking} proposed an algorithm to automatically identify bug-introducing changes with high accuracy by combining the annotation graphs and ignoring non-semantic source code changes. Results show that their algorithm outperforms the \textit{SZZ}. Cathal and Leon~\citep{violation_and_fault} conducted an empirical study to investigate the relation between static warnings and actual faults. More recently, Avgustinov et al.~\citep{Tracking15} proposed to combine several diff-based matching strategies to tackle this problem, which we refer to as the state-of-the-art approach in our study for evaluation and comparison. 

However, a proper examination of the performance of the SOTA approach is still lacking in the field. In this study, we manually crafted a dataset of 3,451 static code warnings and their evolution status from four real-world open-source systems and used it to identify potential improvements in the SOTA approach.

\noindent\textbf{Empirical studies on static bug detectors.}
Researchers have been working on understanding and improving the utilization challenge of static bug detectors. Johnson et al.~\citep{not_use_static} study the reasons that developers do not fully utilize static bug detectors via conducting interviews with developers. Results show developers cannot be satisfied with the current static analysis tools due to the high rate of false positives. This study also provides some suggestions to improve future static tools, e.g., improving the integration of the tool and automatic fixes. Beller et al.~\citep{beller} performed a large-scale study to understand the current status of using static bug detectors in open-source systems, e.g., whether or not used, and what running configurations are used. Wang et al.~\citep{wang2018there} aimed to find whether there is a golden feature to indicate actionable static warnings. Additionally, a survey was conducted by Muske et al.~\citep{muske2016survey} who reviewed static warnings handling studies as well as collected and classified handling approaches.

Studies are also conducted to understand the nature of the issues found by static bug detectors. Ayewah et al.~\citep{google_defects} discuss the defects found by static bug detectors at Google with regards to false positives, types of warnings generated, and their severity. Wedyan et al.~\citep{defect_types} found that the issues by static bug detectors are much more related to refactoring than defects. Habib et al.~\citep{bugs_we_can_find} study the effectiveness of static bug detectors in terms of their ability to find real defects and find that static bug detectors do find a non-trivial portion of defects. An empirical study~\citep{yan2017revisiting} evaluated the degree of correlation between defects and warnings on the evolution of projects. Tomassi et al.~\citep{tomassi2018bugs} examined static bug detectors by considering 320 real Java bugs. Their evaluation shows that static analyzers are not as effective in bug detection, with only one bug detected by {\textit Spotbugs}. Trautsch et al.~\citep{trautsch2019longitudinal} conducted a longitudinal study on static analysis warning trends. They found that the quality of code with regard to static warnings is improving, and the long-term effects of static bug detectors are positive.

Our study focuses on a different aspect, which is to provide better ways to track how static code warnings evolve. Also, our study includes a manual analysis of a non-trivial dataset of static code warnings for the purpose of improving the tracking precision, which is not covered by prior work.  

\noindent\textbf{Utilizing the tracking of static code warnings.}
Better tracking static code warnings across development history provides many benefits. For example, there has been an increasing interest in concluding fix patterns. Kui et al.~\citep{Fixpattern17} mine the fix patterns on static code warnings from the software repository, and the SOTA approach was applied in their research. However, they did not conduct an evaluation on the approach about how accurate the SOTA approach performs. A study~\citep{phoenix} proposed a novel solution to automatically generate code fixing patches for static code warnings via learning from fixing examples. Another recent work~\citep{SAST} proposed a tool to help developers better utilize static bug detectors on security issues by clustering based on common preferred fix locations. This line of work can definitely benefit from an improved tracking approach.
In addition, there have been many works to prioritize and recommend certain types of warnings based on development history. Among them, a study~\citep{fix_warning_first} observed the static warnings in different static bug detection tools and proposed a history-based warnings prioritization to mining the fix cases recorded in the code change history. Results show that over 90\% of warnings remain in the projects or are removed during code non-fix changes. Ted et al.~\citep{rank_warnings} explored the ranking of warnings from static bug detectors and presented a technique with a statistical model to rank the static warnings that are most likely to be true positives. In addition, another work, Quinn et al.~\citep{quinn}, aimed at actionable static warnings, and presented an actionable alert prediction model by creating feature vectors based on code characteristics. In comparison,  our work focuses on the status changes of static warnings in the evolution of software projects. The other work~\citep{burhandenny2016examination} statistically investigated the trend of static warnings over the releases of OSS products and introduced a novel metric (e.g., the index of programmers' attention) to analyze the automatically pointed static warnings and the actual attention that programmers paid to those static warning. Higo et al.~\citep{higo2020ammonia} proposed an approach based on static analysis across the development history to identify project-specific bug patterns. A better tracking mechanism will provide more accurate results for such work.
\section{Conclusions}
\label{sec:conclusions}
Tracking the evolution of static code warnings across software development history becomes a vital question due to the increasing interest in further utilizing static bug detectors by integrating them into developers' workflow. Also, such tracking is widely used in many downstream software engineering tasks.   

This study presents a careful investigation of the performance of the state-of-the-art approach in tracking static code warnings. In particular, a dataset of 3,451 static code warnings for four open-source projects and their evolution status is crafted through manual labeling. Further, we summarize the six causes that introduce false positives for the SOTA approach. To address the false positives, this paper presents an improved approach \textbf{StaticTracker}.  

Last, we perform comparative and generalization evaluations. Results show that our improved approach outperforms the SOTA approach significantly (i.e., \minor{the tracking precision from 64.4\% to 90.3\% ).}

\ifCLASSOPTIONcaptionsoff
  \newpage
\fi



\bibliographystyle{ieeetran}
\bibliography{paper}


\begin{IEEEbiography}[{\includegraphics[width=1in,height=1.25in,clip,keepaspectratio]{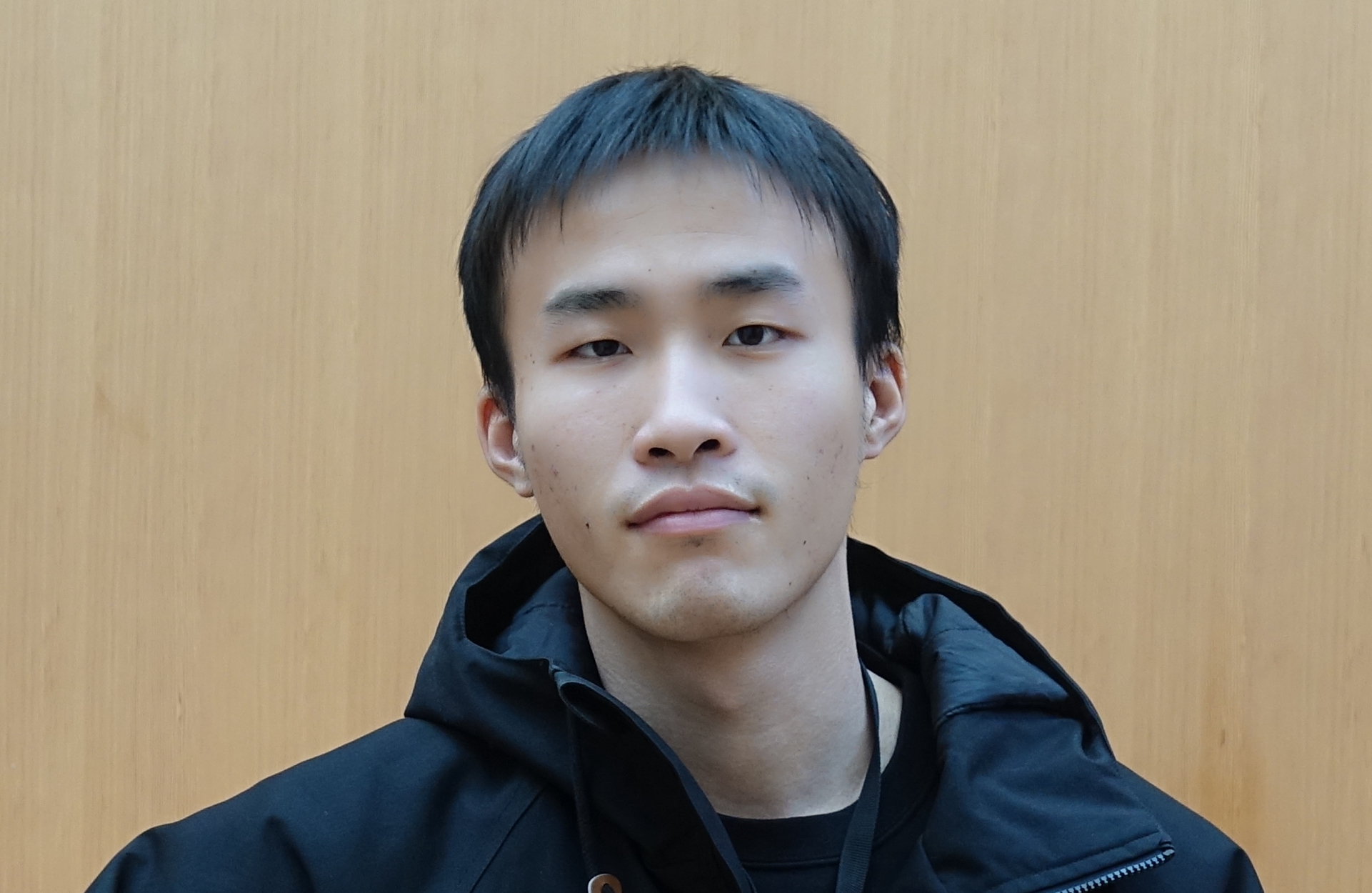}}]%
{Junjie Li}
is a PhD student in the Department of Computer Science and Software Engineering at Concordia University, Montreal, Canada. He obtained MSc. in Computer Science at Concordia University, and received his BSc. in Computer Science of Sichuan University. Contact him at l\_unjie@encs.concordia.ca.
\end{IEEEbiography}

\begin{IEEEbiography}[{\includegraphics[width=1in,height=1.25in,clip,keepaspectratio]{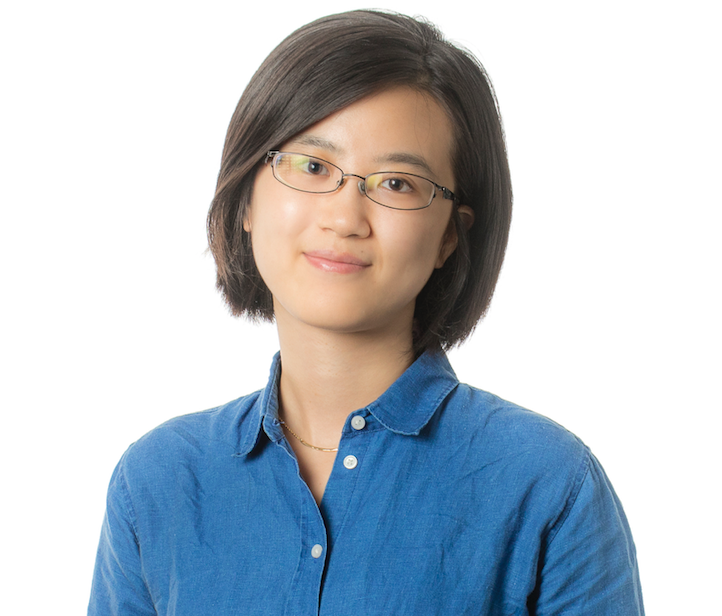}}]%
{Jinqiu Yang}
is an Assistant Professor in the Department of Computer Science and Software
Engineering at Concordia University, Montreal, Canada. Her research interests include automated program repair, software testing, quality assurance of machine learning software, and mining software repositories. Her work has been
published in flagship conferences and journals such as ICSE, FSE, EMSE. She serves regularly
as a program committee member of international conferences in Software Engineering, such as
ASE, ICSE, ICSME and SANER. She is a regular reviewer for Software Engineering journals such as EMSE, TSE, TOSEM and JSS. Dr. Yang obtained her BEng from Nanjing University, and MSc and PhD from University of Waterloo. More information at: https://jinqiuyang.github.io/.
\end{IEEEbiography}

%

\end{document}